\documentclass[a4paper,11pt]{article}
\pdfoutput=1 

\usepackage{jcappub} 

\usepackage[T1]{fontenc} 
\usepackage{graphicx}
\usepackage{amsmath}
\usepackage{upgreek}
\usepackage{float}
\usepackage[caption =  false]{subfig}
\usepackage{graphics}
\usepackage{comment}
\usepackage{bm}
\usepackage{journals}
\usepackage{adjustbox}
\usepackage{natbib}

\title{\boldmath Beyond Best-Fits and Model Selection - Introducing ‘Reliability’ of cusp-core inference of dark matter halos}

\author[a,b]{Manush Manju}
\author[b]{Subhabrata Majumdar}

\affiliation[a]{UM-DAE Center for Excellence in Basic Science, Mumbai, India}
\affiliation[b]{Tata Institute of Fundamental Research, 1 Homi Bhabha Road, Mumbai, 400005, India}

\emailAdd{manush@cbs.ac.in}
\emailAdd{subha@tifr.res.in}

\newcommand{\m}[1]{\mathrm{#1}}

\newcommand{\lm}{\m{log}(M_{200})}
\newcommand{\msun}{\m{M_{\odot}}}
\newcommand{\lms}{\m{log}(M_{200}/\msun)}
\newcommand{\lc}{\m{log}(c_{200})}
\newcommand{\lrhob}{\m{log}(\rho_b)}
\newcommand{\lrhobs}{\m{log}(\rho_b\,/\,[\m{\msun\,kpc^{-3}}])}

\newcommand{\yd}{\Upsilon_{\m{disk}}}
\newcommand{\yb}{\Upsilon_{\m{bul}}}
\newcommand{\mstar}{M_{\m{star}}}
\newcommand{\lmstar}{\m{log}(\mstar/\m{M_{\odot}})}
\newcommand{\mgas}{M_{\m{gas}}}

\newcommand{\lsig}{\m{log}(\Sigma_0)}
\newcommand{\lsigs}{\m{log}(\Sigma_0/[\m{M_{\odot}\,pc^{-2}}])}
\newcommand{\HI}{H\,{\footnotesize{I}}}

\newcommand{\rsize}{r_{\m{size}}}
\newcommand{\rfid}{r_{s,\m{fid}}}
\newcommand{\radii}{\bar{r}}
\newcommand{\rlast}{r_{\m{last}}}

\newcommand{\vstar}{V_{\m{star}}}
\newcommand{\vin}{V_{\m{2kpc}}}
\newcommand{\vmax}{V_{\m{max}}}
\newcommand{\vlast}{V_{\m{last}}}

\newcommand{\Fi}{\m{F}(\radii_i)}
\newcommand{\grid}{(M_{200},\rsize)}
\newcommand{\ori}[1]{\overline{#1}}
\newcommand{\origrid}{(\ori{M_{200}},\ori{\rsize})}
\newcommand{\ipm}{{\mathcal{M}}}

\newcommand{\rela}{\omega_i}
\newcommand{\Relcc}{\mathcal{R}_{\m{c-c}}}
\newcommand{\Relhp}{\mathcal{R}_{\m{h-p}}}
\newcommand{\Relcp}{\mathcal{R}_{\m{c-p}}}
\newcommand{\Rel}{\mathcal{R}}

\newcommand{\chisq}{\chi^2}
\newcommand{\lbf}{\m{ln}\mathcal{B}}
\newcommand{\pres}[1]{\Delta #1/#1}
\newcommand{\accu}[1]{|#1 - \ori{#1}|/\Delta #1}
\newcommand{\mean}[1]{\langle #1 \rangle}

\abstract{Simulated dark matter (DM) halos, across many orders in masses, are amazingly self-similar in their density profiles and well described by the NFW profile over the entire radial range.  Observationally driven departures from the NFW profile in the innermost regions has led to a division between the observed `cores' in lieu of the `cuspy' density profiles found in simulations. The proposed solutions to this so-called \textit{cusp-core} problem include mechanisms ranging from baryonic feedback processes to DM particle physics properties. Even with high-quality observed rotation curves (RCs), it is often difficult to distinguish cusps-vs-cores, and especially one cored profile from another, making inferences coming out of these studies inconclusive. To date, inferences on cusps-vs-cores rely heavily on statistical techniques for model selection and parameter estimation.

In this work, we introduce the notion of a Bayesian analysis motivated `reliability'  that gives a truer distinction of cusp-core and other halo-parameters (like mass-concentration) in an ensemble of observed galaxies. We create hundreds of thousands of realistic mock SPARC RCs,  with both cuspy and cored DM density profiles as model inputs. These RCs carefully incorporate the details of SPARC data such as the nature of observed uncertainties and different sources of scatters arising from observation, presence of baryons, DM mass-concentration, etc. Bayesian analysis of these mock RCs enables us to reconstruct and identify the parameter space in galaxy observable and theory where one can venture beyond best-fits to a preferred DM halo model or model selections between different density models.  We find that its imperative to choose low stellar surface density ($\Sigma_{\star}$) galaxies for reliable cusp-vs-core distinction; for example, RC data for galaxies with $\Sigma_{\star} \leq 2.5$ is needed for a 75\% confidence in distinguishing cusps from cores. Similarly, we also find that for correct estimations of the halo masses and concentrations, the RCs need to be measured to at least a radial distance $\geq 0.8r_s$ where $r_s$ is the scale radii of the corresponding DM halo  density profiles.  Out of the total $\sim$ 135 SPARC galaxies, using our reliability criteria, we find that only 21 RCs clear the bar to be used for any unbiased cusp-core distinction as well as DM halo mass-concentration estimates at  $\geq$75\% reliability confidence level. With  $\geq$66\% ( $\geq$50\%) reliability settings, the sample size increases to 44 (59). Interestingly, in the $\geq 75$\% reliable subsample, there are 5 times more galaxies that are reliably cored than cuspy. We urge the community to use these more reliable `gold' samples in future cusp-core and other DM halo properties inferences.}

\keywords{Bayesian reasoning, rotation curves of galaxies, dark matter theory}

\begin{document}
\maketitle
\flushbottom

\section{Introduction}
\label{sec:Intro}
It has been clear for many decades that the rotation curves (RCs) of disk galaxies point to the presence of significant amounts of dark matter (DM) in their outer regions, where the rotation velocity remains nearly constant ($V_c \sim r^0$), contrary to the expected Keplerian fall ($V_c \sim r^{-1/2}$) \citep{Rubin_1978,Kent_1986}. In the inner regions, a solid-body-like behavior for the DM halo was evident, as the rotation velocity increased roughly linearly with the radius ($V_c \sim r$). These observations indicated a double power law model for the density distribution of DM halos (where $\rho \sim v^2r^{-2}$), with an inner behavior characterized by $\rho \sim r^\alpha$ where $\alpha \approx 0$, and $\rho \sim r^\gamma$ for the outer part with the logarithmic slope, $\gamma \approx -2,$ as suggested by flat RCs \citep{Begeman_1991,kormendy_2004}. An inner slope of $\alpha = 0$ signified the presence of a central core in the DM distribution, spanning a significant portion of the optical disk. As a result, cored models with asymptotic isothermal behavior ($\gamma = -2$), such as \textit{pseudoisothermal} profiles, were recommended as the appropriate models for describing DM in disk galaxies \citep{kormendy_2004}.

 When the results of numerical N-body simulations based on collision-less cold DM (CDM) models became available in the 1990s, a discrepancy was quickly realized: the inner regions of CDM halos showed steeper slopes of $\alpha \approx -1$ in contrast to the observed core-like ($\alpha \approx 0$) behaviors \citep{Navarro_1996}. Among the pioneering works, \citet{Navarro_1997} systematically analyzed CDM halos from their simulation based on various cosmological settings and found that the density profile followed a universal double power law profile, with $\alpha = -1$ and $\gamma = -3$ (hereafter, NFW profile). The study of \citet{Navarro_1997} was strengthened by subsequent simulations with increasing resolutions that predicted steeper profiles in the inner regions of CDM halos in the range, $-0.8 \lesssim \alpha \lesssim -1.5$, where the exact value of $\alpha$ depends on the details of the simulation and the analytical fitting function used to describe the numerically obtained distribution \citep{Colin_2004,Navarro_2004,Merritt_2005}. 

 The observational studies of the central behavior of galactic halos improved as the numerical evidence for the central ``cusp'' in simulated CDM halos continued to grow \citep{Navarro_2004,Graham_2006}. To better understand the distribution of DM in the center, the focus was turned to the RCs of gas-rich dwarf galaxies, whose dynamics are dominated by DM \citep{deBlok_2001}. The mismatch between the simulated cusps and the observed cores was exacerbated by early analyses of the \HI{} RCs of such galaxies \citep{Moore_1994,Flores_1994}. A cuspy profile with $\alpha = -1$ has a RC that rises as $V_c \sim \sqrt{r}$, whereas that of the cores with constant density rises linearly, $V_c \sim r$. The inner shapes of the RCs in the two scenarios are observably diverse and profoundly different. As the evidence for this discrepancy accumulated, the search for a solution to this ``cusp/core'' controversy became a major challenge linking the fields of astrophysics, particle physics, and cosmology \citep{deBlok_2009,Bullock_2017}. 

 Many solutions have since been proposed to resolve the cusp/core problem together with other small-scale problems in both theoretical and observational fronts \citep{Bullock_2017}. The success of $\Lambda$CDM cosmology in explaining the large-scale structure formation makes this `apparent' discordance a puzzling problem. For the same reasons, the proposed solutions aim to keep the large-scale successes of the $\Lambda$CDM model intact and predict alternatives/extensions to CDM that includes a process that effectively transforms the cuspy profile as predicted by CDM to a cored one. Two most extensively studied hypotheses for achieving cusp-to-core transformations in dwarf galaxies include the astrophysical baryonic feedback process and the Self-Interacting DM scenario \citep{Pontzen_2012, Spergel_2000}.

One of the earliest and most prominent hypotheses predicted to address the cusp-core problem relates to the influence of baryonic physics on the structure of CDM haloes \citep{Navarro_1996b}. Baryonic processes, such as supernova explosions (SNe), play a significant role in this context by expelling substantial amounts of gas from the central star-rich regions. This results in a relaxation of the gravitational potential in the inner region, leading to the expansion of DM particles in orbit at the center to higher orbits. It has been demonstrated through semi-analytical calculations and simulations that the spontaneously expanding hot gas bubble introduces sudden perturbations to the gravitational potential, which can disturb the DM component \citep{Pontzen_2012}. Subsequently, the radiative cooling required to lower the gas temperature occurs essentially adiabatically, resulting in an unchanged DM distribution. Such repeated outflows tend to accumulate and gradually push DM away from the center, forming a core in the central region \citep{Pontzen_2012,DiCintio_2014a}.


The Self-Interacting DM (SIDM) hypothesis, initially put forward by \citet{Spergel_2000} as a solution to the cusp-core problem proposes DM to have a self-scattering cross-section that allows it to elastically collide with each other. Consequently, faster DM particles scatter the slower, inner DM particles to higher orbits. These collisions lead to the thermalization of the inner halo, transforming the initial cuspy DM density profile into an isothermal core profile. While several other hypotheses have been proposed to address the cusp-core problem (see for eg., \citep{Marsh_2015,Ferreira_2021,khoury_2015,McGaugh_2016}), the two most extensively studied solutions in terms of density profiles derived from RCs are CDM with SNe feedback and SIDM. Both SNe feedback and SIDM advocate for a core density profile in the inner regions of disk galaxies, albeit with distinct core properties in each case.

SNe feedback models are most effective in large mass galaxies that host a sufficient number of stars capable of bursty blow-out processes. On the other hand, the creation of cores within larger halos becomes increasingly more challenging due to the deep potential wells present in such haloes. Therefore, the most efficient region for cusp-core formation will depend on the interplay between the energy available from SNe explosions and the depth of the halo's potential well \citep{Tollet_2016}. In their cosmological hydrodynamical N-body simulations, \citet{DiCintio_2014a} established a strong correlation between the inner slope of DM haloes resulting from SNe feedback and the stellar-to-halo mass ratio ($\mstar/M_{200}$). In particular, the cusp-core transformation for SNe feedback model is expected to be efficient in the halo mass range, $M_{200}/\msun \in (9.5,\,12)$. The resulting DM density profile of the cored halo is represented by the DC14 profile, characterized by its central slope which depends on $\mstar/M_{200}$.

Core formation due to SIDM is also influenced by the halo potential, owing to the velocity-dependent nature of the self-interaction cross-section \citep{Kaplinghat_2016, Rocha_2013}.  Larger halos accommodate fast-moving DM particles, which results in a weaker efficiency for core formation through DM scatterings. Moreover, for a given halo mass, the distribution of its baryonic components influence both the central density ($\rho_0$), and the core size ($r_c$) \citep{Kaplinghat_2014}. Galaxies with higher baryonic densities play a significant role in shaping the central gravitational potential, giving rise to dense and small cores. In contrast, galaxies with a relatively lower baryonic contribution to their potential lead to cores with lower density and larger values of $r_c$. These characteristics are integrated in the resultant halo by modeling the density profile of SIDM as an isothermal profile in the inner region, where self-interaction dominates, seamlessly transitioning to an outer NFW profile where DM is essentially CDM \citep{Kamada_2017}.

SPARC (\textit{Spitzer} Photometry and Accurate Rotation Curves) comprises a database of nearly 175 disk galaxies with high-quality RCs measured using state-of-the-art techniques \citep{Lelli_2016}.  SPARC galaxies span a wide range of stellar properties and masses ($6 \lesssim \mstar/\msun \lesssim 11$). The central behaviour of SPARC RCs span a diverse behaviour from extremely cuspy to overly cored in nature \citep{Ren_2019}. Over the past half-decade, the SPARC RC sample was the focal point for various DM studies, including the cusp-core problem, DM halo density profiles, and other statistical analyses related to mass modeling of RCs. DM density profiles proposed for both SNe feedback and SIDM have been demonstrated to provide excellent fits to this RC dataset. Additionally, the diverse behavior of the inner regions of RCs was found to be efficiently addressed by these profiles \citep{Creasey_2016,Santos_2020}. It was also noted that low surface brightness galaxies statistically exhibit a preference for cored profiles over cuspy or NFW profiles \citep{Relatores_2019,Li_2020,Frosst_2022}. Nevertheless, a substantial fraction of these RCs are equally or better fitted by NFW profiles compared to cored profiles.

The assessment of the goodness of fit by an assumed DM density profile is commonly carried out using the reduced $\chi^2$ and/or the cumulative distribution function (CDF) of the reduced $\chi^2$s for all RCs \citep{Ren_2019,Katz_2017}. As has been observed in recent works, this method of model comparison is found to be flawed when dealing with non-linear fitting models \citep{Cameron_2020,Zentner_2022}. In fact, $\chi^2$ cannot be considered a reliable indicator of how effectively a model accounts for observed data, particularly when RCs are frequently over-fitted. This highlights the necessity to address a more fundamental question: Are cores present in disk galaxies? This question takes precedence over the problem of testing hypotheses for cusp-core transformation using RCs. Considering the gravity of the cusp-core issue, there is an urgent need for a more robust and suitable technique to distinguish cores from cusps in observed RCs.

Distinguishing one cusp-to-core transformation hypothesis from another through the modeling of RCs has proven to be ineffective, as both the cored profiles in SIDM and SNe feedback models exhibit an equal degree of compatibility with the RC data \citep{Zentner_2022}. In such instances, identifying the original mechanism responsible for cusp-core transformation necessitates moving beyond the analysis of goodness-of-fit and considering the characteristics of the mechanism and its agreement with the overall observations. This involves finding the range of galaxies with cored DM density profiles and evaluating how well a specific hypothesis can produce cores in those galaxies. Furthermore, it entails an examination of the general properties of observed cores and halo parameters. Achieving this relies on the reliability of model selection and parameter estimation techniques to accurately classify a RC as either cuspy or cored, and to extract its halo (and core) parameters.

In this work, thousands of mock RCs closely resembling those in the SPARC dataset were generated. The scatter, error bars, and the distribution of radial data points for these mock RCs was carefully incorporated, following the observed trends in SPARC RCs. Both cuspy and cored RCs were constructed, assuming an NFW profile and a model-independent cored profile, respectively. These RCs were developed within a two-dimensional parameter space, denoted as $(M_{200}, \rsize)$, where $\rsize = r_{\m{last}}/r_s$ represents the observed size or extent of the RC (radius of the last data point) in units of $r_s$, the scale radius of the DM (NFW) density profile, drawn from the predicted cosmological mass-concentration scaling relation. Baryonic components, including gas and stellar disks, were incorporated based on observed scaling relations and abundance matching relations, accounting for the inherent variations in disk scale lengths and total masses. Subsequently, these mock RCs were analyzed in a manner akin to the observed RCs. The fit results underwent rigorous model selection and other statistical criteria to identify the most suitable model, which was further cross-validated against the original model used in its creation.

The paper is structured as follows. In Section~\ref{sec:sparc}, our fitting procedure and the results obtained from its application to the observed SPARC RCs are presented. Section~\ref{sec:mock} provides an overview of the mock RCs and their characteristics derived from SPARC. The analysis of mock RCs to assess the reliability of RC modeling is discussed in Section~\ref{sec:rel}. This information is employed to identify the most reliable RCs in the SPARC database, and the corresponding results are presented in Section~\ref{sec:gal}. Finally, the main findings are summarized in Section~\ref{sec:con}.

\section{Observed RCs}
\label{sec:sparc}
In this section, we summarize the details of observed RCs used in this study, and the bayesian analysis done on it to study the DM halo properties. 
\subsection{SPARC RCs}
SPARC database consists of high-quality HI/H$\alpha$ RCs of 175 late-type galaxies spanning a broad range of morphologies (S0 to Irr) \citep{Lelli_2016}. The stellar disk mass distributions of SPARC data is traced using spitzer photometry at 3.6$\mu$m, which falls in the near-infrared band that is apt for breaking the disk-halo degeneracy \citep{vanAlbada_1985}. 135 out of the 175 galaxies have extended RC data spanning the flat part, for which the velocity of the flat region, $V_{\rm{flat}}$, is available. 

SPARC provides the total circular velocities ($V_c$) at galactocentric radii ($r$), which can be decomposed into contributions from various components of the galaxy as,
\begin{equation}
    \label{eq:s1}
    V^2_c(r) = V_{\rm{dark}}^2(r) + |V_{\rm{gas}}|V_{\rm{gas}}(r)+ \Upsilon_{\rm{disk}}V_{\rm{disk}}^2(r) + \Upsilon_{\rm{bul}}V_{\rm{bul}}^2(r),
\end{equation}
where $V_{\rm{dark}}$ is due to the DM halo, which we assume throughout as spherically symmetrical. $V_{\rm{gas}}$ is the contribution from gas disk (which we assume to be a thin exponential disk), $V_{\rm{disk}}$ is the contribution from stellar disk (also assumed to be a thin exponential disk) and $V_{\rm{bul}}$ is the contribution from stellar bulge\footnote{The bulge distribution was calculated assuming a spherical profile in SPARC, but we use an exponential disk to fit $V_{\rm{bul}}$ and finds it also a good description of the given data.}. Only 32 galaxies have non-zero bulge components. The absolute value of $V_{\rm{gas}}$ is required because $V_{\rm{gas}}$ can sometimes be negative, due to significant central depressions resulting in an overall net outward gravitational force.

$\yd$ and $\yb$ are the stellar mass-to-light ratios of disk and bulge components respectively. SPARC provides $V_{\rm{gas}}$, $V_{\rm{disk}}$ and $V_{\rm{bul}}$ data of each galaxy at $r$ where $V_c$ is also available. For a given $\yd$ and $\yb$, we fitted each of the baryonic components, $V_{\rm{gas}}$, $V_{\rm{disk}}$ and $V_{\rm{bul}}$, assuming exponential disk profiles to obtain the mass and radial scale of the gas disk, stellar disk, and bulge ($(M_{\rm{gas}}, R_{\rm{gas}}$) $(M_{\rm{disk}}, R_{\rm{disk}})$ and $(M_{\rm{bul}}, R_{\rm{bul}})$, respectively).

\subsection{DM density profiles}
The DM contribution to the RC can be obtained by subtracting the baryonic contribution from the total rotation velocity. Given a radial density profile for DM distribution ($\rho_{\rm{dark}}$), the DM contribution to total rotation velocity is,
\begin{equation}
    \label{eq:s2}
    V_{\rm{dark}}(r) = \sqrt{\frac{GM_{\rm{dark}}(r)}{r}};\,\, M_{\rm{dark}}(r) = 4\pi\int_{0}^{r}\rho_{\rm{dark}}(r')r'^2\mathrm{d}r'.
\end{equation}

To model the cuspy density profile ($\rho_{\rm{cusp}}$) associated with the CDM assumption, we employ the well-established NFW profile \citep{Navarro_1997},
\begin{equation}
    \label{eq:s3}
     \rho_{\rm{NFW}}(r) = \frac{\rho_s}{\bigg(\cfrac{r}{r_s}\bigg)\bigg[1 + \bigg(\cfrac{r}{r_s}\bigg)\bigg]^2}.
\end{equation}
The NFW profile is characterized by two free parameters: the density scale, $\rho_s$, and the radial scale, $r_s$.  These parameters allow for the determination of virial quantities, including $R_{\Delta}$, which signifies the virial radius where the mean halo density reaches $\Delta$ times the critical density of the Universe, and $M_{\Delta}$, the mass enclosed within $R_{\Delta}$. Throughout this study, we adopt the assumption that $\Delta = 200$ and work with the corresponding virial quantities, denoted as $R_{200}$ and $M_{200}$. The halo concentration is then defined as,
\begin{equation}
    \label{eq:s4}
    c_{200} = \frac{R_{200}}{r_s}.
\end{equation}
Instead of $(\rho_s, r_s)$ we set $(\rm{log}(M_{200}),c_{200})$ as the two free parameters that define an NFW halo and consider them as the free parameter for fitting the NFW velocity profile to the data.

To model the galactic RCs in which a cusp transitioned to a core, we employ a generic cored profile that is independent of any specific hypotheses for core formation processes. Taking inspiration from models like coreNFW \citep{Read_2016} and SIDM \citep{Kaplinghat_2016}, we consider a three-parameter DM profile, defined as,
\begin{equation}
    \label{eq:s5}
    \rho_{\rm{core}}(r) =  
    \begin{cases}
        \rho_{\rm{burk}}(r),    & \text{for } r \leq r_c\\
        \rho_{\rm{NFW}}(r),     & \text{for } r > r_c,
    \end{cases}
\end{equation}
where, $\rho_{\rm{burk}}$ represents the phenomenological Burkert profile, proposed by \citet{Burkert_1995}, which was shown to be self-similar and capable of fitting the RCs of dwarf galaxies, after appropriate scaling. This profile is characterized by $\alpha = 0$ and $\gamma = -3$, as defined by:
\begin{equation}
    \label{eq:s6}
    \rho_{\rm{burk}}(r) = \frac{\rho_b}{\bigg(1+\cfrac{r}{r_b}\bigg)\bigg[1+\bigg(\cfrac{r}{r_b}\bigg)^2\bigg]},
\end{equation}
where $\rho_b$ and $r_b$ are the central density and the scale radius respectively. We make the assumption that the density and mass profiles at $r = r_c$ are continuous, 
\begin{equation}
    \label{eq:s7}
    \rho_{\rm{burk}}(r_c) = \rho_{\rm{NFW}}(r_c), \text{ and } M_{\rm{burk}}(r < r_c) = M_{\rm{NFW}}(r < r_c).
\end{equation}
With these conditions, the NFW parameters $(\rm{log}(M_{200}), c_{200})$ can be mapped to the Burkert parameters $(\rho_b, r_b)$, making $\rho_{\rm{core}}$ a three parameter profile. $r_c$ denotes the radius beyond which the NFW profile characterizing the CDM halo is retained and signifies the size of the core. Henceforth, $r_c$ will be referred to as the core size, and $r_b$ as the scale radius of the core.

\subsection{Bayesian analysis on RCs}
\label{ssec:bayes}
A Bayesian analysis was conducted on all 175 RCs in the SPARC database, wherein both the cuspy (eq.~\eqref{eq:s3}) and the cored (eq.~\eqref{eq:s5}) density profiles, along with the baryonic components as described in (eq.~\eqref{eq:s1}), are fitted to the RCs. The posterior distributions of the DM parameters, as well as $\yd$ (and $\yb$), were derived using the multimodal nested sampling algorithm implemented in the open-source Python package UltraNest \citep{Buchner_2019}.

UltraNest provides an efficient solution for determining the best-fit models from multimodal posterior distributions and addressing challenges related to significant parameter degeneracies, often encountered when fitting DM halo parameters, including disk-halo degeneracies, to RCs. In contrast to traditional Markov Chain Monte Carlo (MCMC) methods, the nested sampling algorithm primarily computes the Bayesian evidence value, $Z_{H}$, as the final outcome, where $H$ represents the hypothesis under investigation, while also providing posterior probabilities as a supplementary result \citep{Skilling_2004}. The importance of these evidence values lies in their utility for calculating the Bayes Factor, which serves as a powerful tool for model selection.

The likelihood function used in this study was,
\begin{equation}
    \label{eq:s8}
    \mathrm{ln}\mathcal{L}(\bm{\uptheta}) = \frac{1}{2}\sum_{r}\frac{(V_{c,\rm{obs}}(r) - V_{c,\rm{model}}(r,\bm{\uptheta}))^2}{\Delta V^2_{c,\rm{obs}}(r)},
\end{equation}
where $\bm{\uptheta} = (\bm{\uptheta}_{\rm{D}},\yd,\yb)$ are the parameters of the DM profiles and the mass-to-light ratios of the stellar and bulge components which are considered for fitting. $V_{c,\rm{obs}}$ represents the observed RC data, $V_{c,\rm{model}}$ denotes the total rotation velocity dependent on $\bm{\uptheta}$, and $\Delta V_c$ stands for the quoted error associated with the total circular velocities. The summation is performed over all available data points $r$.

For the baryonic parameters ($\yd$ and $\yb$) we used log-normal priors centered at 0.5 and 0.7, respectively, with a standard deviation of 0.1 dex, as suggested by stellar population synthesis models \citep{Schombert_2019}. Additionally, a constraint of $M_{\rm{gas}}+M_{\rm{disk}}+M_{\rm{bul}} \leq 0.2M_{200}$ was applied to ensure that the baryonic fraction remains consistently below the cosmological value, as suggested by \citep{Katz_2017}.

\begin{table}
	\centering
	\caption{Details of the prior distributions employed for different parameters within the flat and $\Lambda$CDM prior sets. Here, $M_{\m{bary}}$ denotes the total baryonic mass, defined as $M_{\rm{gas}}+M_{\rm{star}}$ where $M_{\rm{star}} = M_{\rm{disk}}+M_{\rm{bul}}$. The uniform distribution ranges are indicated within square brackets, while log-normal distributions are specified with the mean $\pm$ scatter.}
	\label{tab:t1}
	\begin{tabular}{|l|c|r|} 
		\hline
		Parameter  & Prior set & Distribution\\
		\hline
        log$(M_{200}/\rm{M_{\odot}})$ & All & [7,14]\\
        $c_{200}$ & Flat & [1,100] \\
                  & $\Lambda$CDM & MCR $\pm$ 0.11 dex \\
        $r_c$ (kpc) & All & [0.1,$r_s$] \\
		$\yd (\m{M_{\odot}/L_{\odot}})$  & All & log(0.5) $\pm$ 0.1 dex\\
		$\Upsilon_{\rm{bul}}$ ($\rm{M_{\odot}/L_{\odot}}$) & All & log(0.7) $\pm$ 0.1 dex\\
		$M_{\rm{bary}}/M_{200}$ & All & [0,0.2] \\
        $\cfrac{\m{log}(\vmax/V_{\m{flat}})}{\m{log}(2)}$ & All & 0 $\pm 0.5$\\
		\hline
	\end{tabular}
\end{table}

For the DM parameters, two distinct sets of priors were considered, as detailed in Table~\ref{tab:t1}: the flat and $\Lambda$CDM prior sets. In the case of the flat prior set, non-informative uniform priors spanning the most significant physical ranges were considered: $7 \leq \rm{log}(M_{200}/\rm{M_{\odot}}) \leq 14$ and $1 \leq c_{200} \leq 100$. We make the assumption that the core formation processes are not strong enough to create a core of size which exceeds the scale radius of the corresponding CDM halo. This assumption is in general agreement with the simulation results of SIDM \citep{Rocha_2013} and SNe feedback models \citep{DiCintio_2014b}. Thus for the core size, we fix a uniform prior of $0.1 \leq r_c/\mathrm{kpc} \leq r_s(M_{200},c_{200})$.

The virial mass and concentration of CDM haloes are known to follow a power-law relationship \citep{Maccio_2008}, expressed as:
\begin{equation}
\label{eq:s9}
\mathrm{log}c_{200} = a - b \, \mathrm{log}(M_{200}/[10^{12}h^{-1}\rm{M_{\odot}}]),
\end{equation}
The specific values of $a$ and $b$ vary depending on the assumed cosmology. In this study, we adopt the values provided by \citep{Dutton_2014}, with $a = 0.905$ and $b = 0.101$, accompanied by an intrinsic scatter of 0.11 dex. It is important to note that the distances for certain galaxies within the SPARC database were calculated using a Hubble constant of $H_0 = 73 \, \rm{km \, s^{-1}Mpc^{-1}}$, hence, we employ the value of $h = 0.73$ accordingly. To enhance the precision of our constraints on the CDM halo parameters, we implement a $\Lambda$CDM prior set. In this prior set, the prior distribution for $c_{200}$ is defined as a log-normal distribution, with mean and scatter values derived from the CDM mass-concentration relation (MCR), as expressed in eq.~\eqref{eq:s9}.

In their analysis of the SIDM model applied to SPARC RCs, \citet{Ren_2019} introduced a uniform regulation prior, defined as $1/\sqrt{2} \leq V_{\rm{max}}/V_{\rm{flat}} \leq \sqrt{2}$. This prior was intended to ensure that the maximum circular velocity predicted for the DM component ($V_{\rm{max}}$) aligns with the observed velocity at the outer regions of the RC ($V_{\mathrm{flat}}$). For both the flat and $\Lambda$CDM prior sets, we also incorporate a regularization prior, but in the form of a normal distribution for $\mathrm{log}(\vmax/V_{\mathrm{flat}})/\mathrm{log}(2)$, centered at 0, with a scatter of 0.5 dex. The reason for the requirement of a Gaussian regularization prior is discussed in appendix~\ref{sec:appA} along with the details of various other sets of priors considered to fit the RCs. 

\subsection{Fit results}

We have applied the following sample selection criteria to the SPARC dataset: (i) the RC should contain a minimum of 5 data points, (ii) the galaxy's inclination, denoted as $i$, must be equal to or greater than $30^\circ$. The first criterion ensures compatibility with our DM + baryon fitting model, which includes a maximum of 4 free parameters (or 5 if a bulge is present). The second criterion is implemented to minimize the correction required for the uncertainties in inclination, which affect the observed velocities of the galactic disk on the sky-plane ($\sim 1/\mathrm{sin}(i)$). Additionally, only RCs with recorded $V_{\rm{flat}}$ values were included in the analysis, making them eligible for the application of the regularization prior. These criteria led to a total of 128 galaxies included in our study.

\begin{figure*}
    \centering
	\includegraphics[width=\linewidth]{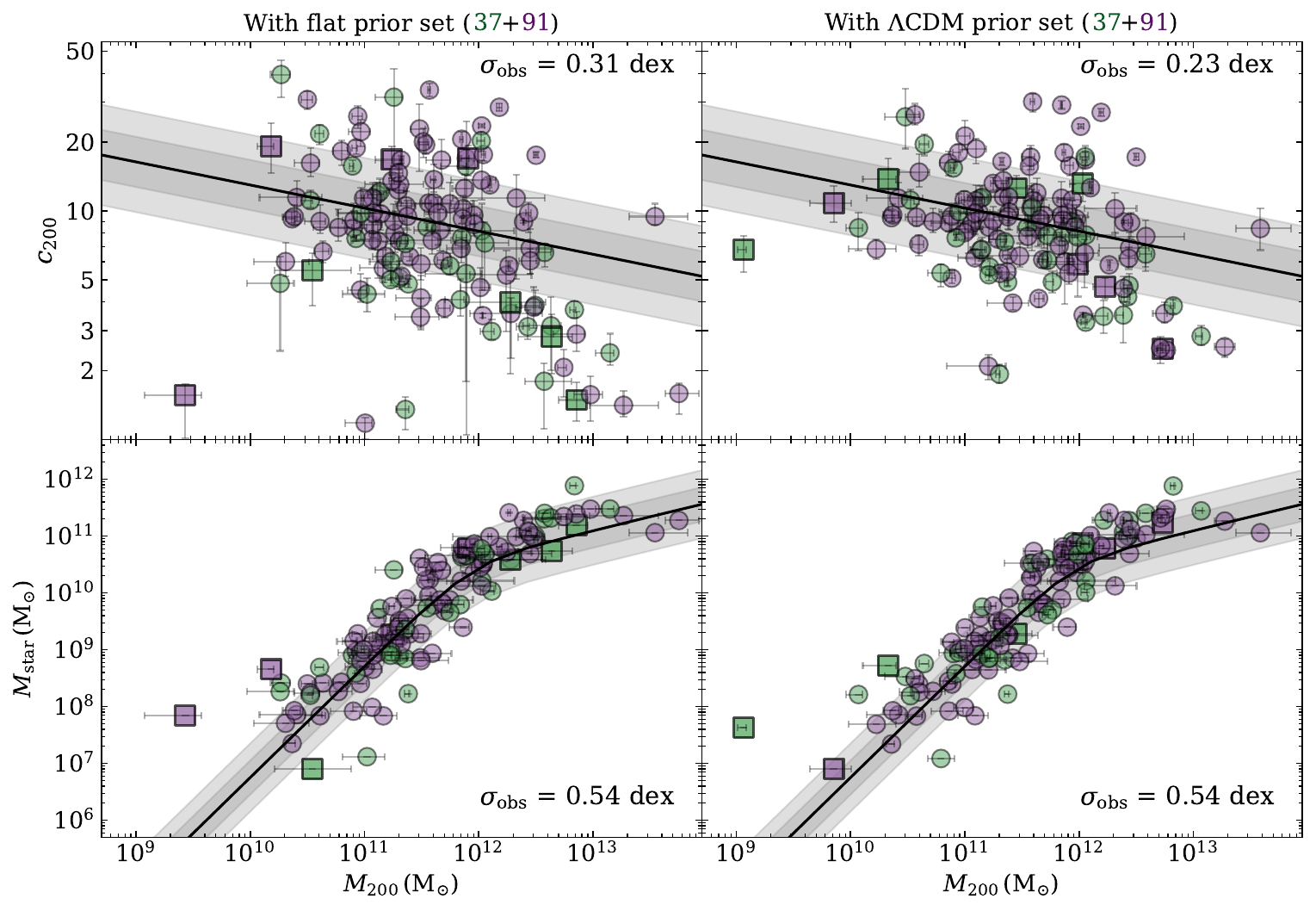}
    \caption{\textit{Top row}:The best-fit values of $M_{200}$ and $c_{200}$ are plotted along with the mean MCR, as given by eq.~\eqref{eq:s9} (black line), together with the predicted 1$\sigma$ (2$\sigma$) scatter represented in dark (light) bands. \textit{Bottom row}: The plots showing the best-fit values of $M_{\rm{star}}$ vs. $M_{200}$. Also shown is the mean stellar-halo mass relation from \citet{Behroozi_2019} (black line), along with its 1$\sigma$ (2$\sigma$) scatter in dark and (light) bands.  The \textit{left column} displays the results obtained from fits with flat prior sets, while the \textit{right column} presents those using the $\Lambda$CDM prior set. The data points are colored in green if the best-fit DM profile is cuspy and in purple if it is cored. In each column, the total number of galaxies is indicated at the top in brackets, denoted as the sum of the green and purple data points. Galaxies for which the best-fit model differs between the columns are marked by squares. The observed scatter around the predicted mean relations ($\sigma_{\m{obs}}$) are provided in each panel.}
    \label{fig:fig_1}
\end{figure*}
We have analyzed the posterior samples derived from the UltraNest fits using the GetDist Monte Carlo Sample analysis Python package \citep{Lewis_2019}. For each fit, we have recorded the mean values from the marginalized posterior distributions as the best-fit parameter values. The lower and upper errors have been computed based on the 68\% confidence band limits. In figure~\ref{fig:fig_1}, we present the fit results, depicting the values of $c_{200}$ and $M_{\rm{star}}$ plotted against $M_{200}$. These results are compared with the expected CDM scaling relations, shown in the top row as the MCR and in the bottom row as the stellar mass-halo mass (SMHM) relation. We have considered two sets of priors for comparison, the flat and the $\Lambda$CDM prior sets (indicated at the top of each column). In the plots, data points are color-coded as green or purple depending on the logarithm of the Bayes factor, defined by:
\begin{equation}
\label{eq:s10}
\mathrm{ln}\mathcal{B} = \mathrm{ln}Z_{\mathrm{core}} - \mathrm{ln}Z_{\mathrm{cusp}},
\end{equation}
where a green point (purple point) corresponds to $\mathrm{ln}\mathcal{B} < 0$ ($\mathrm{ln}\mathcal{B} > 0$).

Bayes factor, $\mathcal{B} = Z_{\mathrm{core}}/Z_{\mathrm{cusp}}$, is a crucial tool in Bayesian analysis for hypothesis testing. It provides the evidence of a statistical model against a null hypothesis, (in this case, the cuspy profile)\citep{Kass_1995}. The Bayesian evidence, denoted as $Z_{H}(\bm{D}|H)$, represents the posterior odds of a hypothesis ($H$) given the data ($\bm{D} \equiv \{D_j\}_{j = 1}^n$), and it is calculated as 
\begin{equation}
    \label{eq:eq11}
    Z_{H}(\bm{D}|H) = \int \mathcal{L}(\bm{\uptheta}|\bm{D})\mathcal{G}(\bm{\uptheta})\mathrm{d}^{d}\bm{\uptheta},
\end{equation}
where, $d$ denotes the dimensionality of the parameter vector space $\bm{\uptheta}$ and $\mathcal{G}(\bm{\uptheta})$ represents the prior distribution of $\bm{\uptheta}$. Consequently, a larger evidence value for the hypothesis $H$ indicates that more regions in the parameter space of $H$ are likely or that a specific region of it is particularly adept at explaining the data \citep{Feroz_2008}. In the context of model selection, the logarithm of the Bayes factor, as given by eq.~\eqref{eq:s10}, can be interpreted as follows: very strong evidence for the core if $\ln\mathcal{B} > 5$, strong evidence for the core if $5 \ge \ln\mathcal{B} > 3$, moderate evidence for the core if $3 \ge \ln\mathcal{B} > 1$, and not worth more than a bare mention if $1 \ge \ln\mathcal{B} > 0$ \citep{Kass_1995}.

In figure~\ref{fig:fig_1}, the number of green points (cusp) and the number of purple points (core) for each prior set are indicated above each column in brackets. Only $\approx$ 28\% of the total galaxies in each column exhibit evidence for a cusp (ln$\mathcal{B} < 0$). Among these, there are 8 galaxies for which the Bayes factor changes sign across the columns, i.e, model selection results varies with the assumed prior sets (square points in figure~\ref{fig:fig_1}). For the remaining galaxies, represented by circle points, the results of model selection are consistent across both sets of priors. Among the 8 galaxies sensitive to the prior, ln$\mathcal{B}$ values fall within the range of -1 to 1, indicating evidence for neither cusp nor core, with both prior sets. Excluding those 8 galaxies, approximately 34\% of galaxies have $|\rm{ln}\mathcal{B}| > 5$, 8\% have $5 \geq |\rm{ln}\mathcal{B}| > 3$, 34\% have $3 \geq |\rm{ln}\mathcal{B}| > 1$, and 23\% of galaxies have $1 \geq |\rm{ln}\mathcal{B}| < 0$.

In the flat prior case (column 1), the observed scatter ($\sigma_{\rm{obs}} = 0.31$ dex) significantly exceeds the predicted scatter (0.11 dex) around the mean MCR. In the $\Lambda$CDM prior case, the inclusion of the MCR as a prior improves its adherence to the relation, although it still exhibits a higher scatter than anticipated ($\sigma_{\rm{obs}} = 0.23$ dex). The predicted scatter for the SMHM relation is 0.3 dex, and the results from both prior sets only agree within twice this scatter. It is noteworthy that, without the regularization prior, this scatter was nearly twice as high, underscoring the significance of the criteria applied to $V_{\m{max}}$ in constraining and resolving degeneracies between the DM halo and the baryonic components. Note that, as expected for Gaussian posteriors, the 68\% error bars on fit parameters decrease as stricter priors were incorporated.

\subsection{Baryonic Scaling Relations}

SPARC provided the total \HI{} mass, allowing us to derive the total mass of the gas component using a constant \HI{}/He fraction: $\mgas = 1.33M_{\m{\HI{}}}$ \citep{Lelli_2016}. Subsequently, we fitted a thin exponential disk model to $V_\mathrm{gas}$ to obtain the gas scale radius, $R_{\m{gas}}$. SPARC offered the scale radius of the stellar disk component, $R_{\m{star}}$. Using the best-fit $\yd$ value, we then determined $\mstar$ by fitting a thin exponential disk model to $\sqrt{\yd}\vstar$, with $R_\m{star}$ fixed at the SPARC-provided value. Utilizing these quantities, $\mgas$, $R_{\m{gas}}$, $\mstar$, and $R_{\m{star}}$, we obtained the following scaling relations:,\begin{align}
    \label{eq:eq12}
    \m{log}(M_{\m{gas}}) &= 0.52\,\m{log}(M_{\m{star}}) + 4.44\,: \,\, \sigma_{\m{obs}} = \pm 0.33 \text{ dex}, \nonumber \\
    \m{log}(R_{\m{gas}}) &= 0.59\,\m{log}(M_{\m{gas}})\, - 4.80\,: \,\, \sigma_{\m{obs}} = \pm 0.15 \text{ dex}, \\
    \m{log}(R_{\m{star}}) &= 0.92\,\m{log}(R_{\m{gas}})\,\, - 0.38\,:  \,\, \sigma_{\m{obs}} = \pm 0.24\,\text{ dex}. \nonumber
\end{align}

\begin{figure*}
    \centering
    \includegraphics[width = \linewidth]{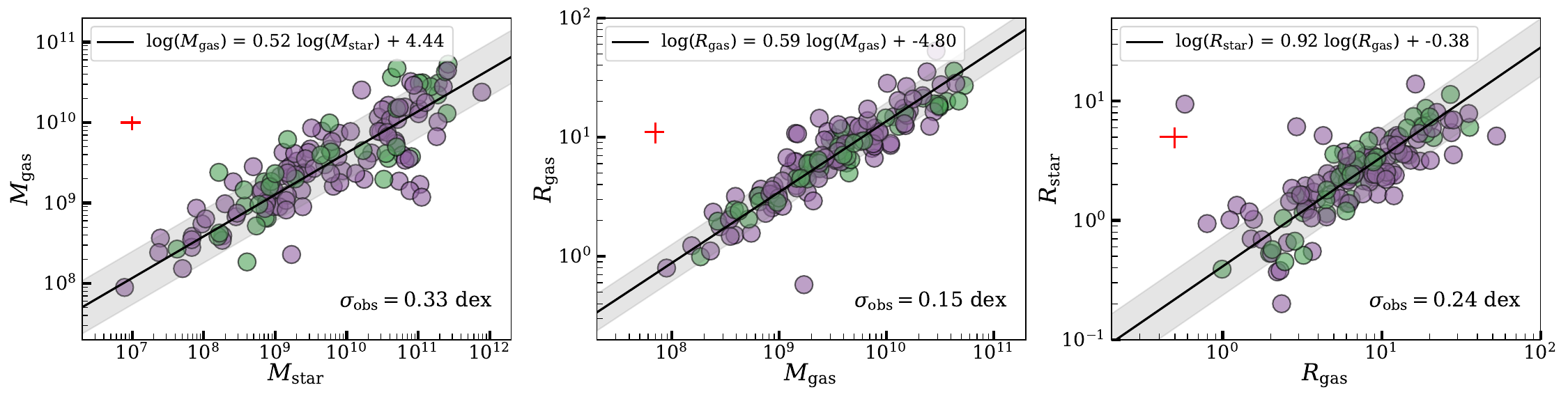}
    \caption{Scaling relations of baryonic components from SPARC RC fits using the $\Lambda$CDM prior set. \textit{Left panel}: $M_\m{star}$ vs $M_\m{gas}$, \textit{middle panel}: $M_\m{gas}$ vs $R_{\m{gas}}$, and \textit{right panel}: $R_{\m{gas}}$ vs $R_{\m{star}}$. The green (purple) points represent galaxies that were found to have a cuspy (cored) profile as the best-fit model based on the Bayes factor. The black line illustrates the best-fit line for the log$(x)$ vs log$(y)$ plot (indicated in the upper left corner), while the grey band indicates the observed 1$\sigma$ scatter (specified in the lower right corner of each panel). The red plus symbol denotes the typical $2\sigma$ error bars on the data points ascribing a minimum of 10\% error on each quantity \citep{Lelli_2016}.}
    \label{fig:fig_3}
\end{figure*}
The quantities $M_{\m{gas}}$, $R_\m{gas}$, and $R_\m{star}$ remain independent of any fit results (to the total circular velocity); they can be directly derived from their respective components in the RC (with a 10\% error margin included). In contrast, $\mstar$ depends on the fit results, particularly influenced by $\yd$, and varies depending on the DM model assumed for the fit. For the relations presented in eq.\eqref{eq:eq12}, we used the best-fit $\yd$ values obtained for the best model as favored by the value of $\lbf$ within the $\Lambda$CDM prior set. It's worth noting that these relations remain consistent when the results from the flat prior set were considered. We've confirmed that the $M_\m{gas}$-$M_\m{star}$ relation, which relies on the fitted value of $M_\m{star}$, is in line with previously observed relations, such as the one in \citet{Papastergis_2012}. In figure\ref{fig:fig_3}, we illustrate the mean scaling relations and the observed scatter corresponding to eq.~\eqref{eq:eq12}.

\section{Mock RCs}
\label{sec:mock}
In this paper, our primary objective is to assess the reliability of inferring cusps and cores by fitting density profiles to velocity RCs. To achieve this, we have generated realistic mock RCs that encompass all the fundamental attributes observed in SPARC RCs.  This includes capturing the mean features and their scatter in the total circular velocity as well as its baryonic components. In this section, we outline our methodology for constructing mock RCs while integrating specific features observed in SPARC RCs.

\subsection{Details for mock RCs from SPARC RCs}
We generated mock RCs in a 2-dimensional parameter space defined by $M_{200}$ and $\rsize$, where $\rsize$ represents the ``observed'' size of a galaxy, defined in units of $r_s$ as
\begin{equation}
    \label{eq:m1}
    \rsize = \frac{r_{\m{last}}}{r_s},
\end{equation}
where $r_\m{last}$ is the radius where last data point is available. The mock RCs were systematically created in a grid format, spanning $15 \times 15$ cells, covering the following ranges: $9 \leq \m{log}(M_{200}/\m{M_{\odot}}) \leq 13.5$ and $0.2 \leq \rsize \leq 5.4$. The grid edges were uniformly distributed within the specified range for $M_{200}$, and for $\rsize$, we used grid-widths of 0.2 when $\rsize \leq 1$ and 0.4 when $\rsize > 1$.

The definition of $\rsize$ is useful in two ways: (i) to create mock RCs of various radial extents; and (ii) to analyze the SPARC RCs to obtain the frequency distribution of the radial data points. For the latter, the radius corresponding to each data point was scaled with a fiducial scale radius, $\rfid$. Fiducial because it will only be conclusively determined at the end of the study whether the $r_s$ obtained from fitting, and the associated model, accurately describes the actual value, if such a determination is even possible. In the bulk of this study, we consider $\rfid$ as the minimum of the $r_s$ values quoted in the SPARC database from their fits to the RCs using the density profiles DC14, coreNFW, and NFW, with $\Lambda$CDM priors \citep{Li_2020}.  Importantly, we have explicitly confirmed that our final results remain consistent even if we opt to use the $r_s$ associated with the best-fit model resulting from our fits to the SPARC RCs.

The scaled radial points ($\radii = r/\rfid$) of all SPARC RCs were then collected into 15 bins (with bin edges $\bar{r}_i$) having bin-widths of 0.2 if $0 < \radii \leq 1$ and 0.4 if $1 < \radii \le 5.4$. The number of radial data points for mock RCs within each scaled radii bin, denoted as $\m{F}(\radii_i)$, was computed as the mean number of observed data points in each bin. Since not all SPARC RCs had data points in every $\radii$ bin, $\m{F}(\radii_i)$ was calculated by taking the mean number of data points separately for each bin, considering only the galaxies with at least one data point in that bin. For a given galaxy, gal, in SPARC, let $\mathcal{R}^i_{\m{gal}} = \{\radii': \radii_i \leq \radii' \leq \radii_{i+1} \: \forall \: \radii' \in \text{RC of gal}\}$, where $\radii_i$ represents the $i^{\m{th}}$ bin edge. Then the radial function of the number of data points in a bin was given by,
\begin{equation*}
    \m{F}(\radii_i) = \frac{\sum\limits_{\m{gal}} \#\mathcal{R}^i_{\m{gal}}}{\#\{\m{gal}: \mathcal{R}^i_{\m{gal}} \neq \upphi \:\: \forall \: \text{gal in SPARC} \}}, 
\end{equation*}
where the symbol $\#$ represents the cardinality of the set, $\upphi$ denotes a null set and the sum is over all galaxies in SPARC database. In a mock RC, the radial points falling in the $i^\m{th}$ bin would have $\m{F}(\radii_i)$ data points distributed uniformly between $\radii_i$ and $\radii_{i+1}.$

Next, with the same radial binning as used above, we considered the distribution of the percentage error of the total circular velocities in each of the 15 bins. The percentage error for each data point in the $i^\m{th}$ bin of a mock RC was then randomly sampled from the corresponding error distributions of SPARC RCs obtained for that bin. For this, we first created the CDFs of the percentage error distributions, and then randomly selected a number from a uniform distribution between 0 and 1. The percentage error of a given data point was then determined as the value of the inverse CDF at that random number. It should be noted that the SPARC database does not provide error bars on the baryonic contributions, so the error bars in the mock RCs were applied only to the total circular velocity.

Finally, we considered the scatter in the total circular velocity of the mock RCs. In the fitting procedure, $\yd$ is the parameter that affects the baryonic components, causing the stellar velocity component to be scaled up or down by a factor of $\sqrt{\yd}$. The DM component we fit to the total RC is a smooth curve. Therefore, the scatter in the total circular velocity should be attributed to the scatter in the stellar and gaseous components. We found that fitting the SPARC galaxies with the given baryonic components and a smooth DM profile did not perfectly describe the RCs. The resulting residuals for both cuspy and cored model fits were often larger than the quoted error bars. To emphasize this scatter, in the mock RCs, we did not consider any scatter for the baryonic components rather we attribute this scatter to the DM component.

The scatter in the total circular velocities of a mock RC is introduced by adding a scatter to the $c_{200}$ parameter employed in generating each specific mock RC at every data point. We estimated that the residuals of the fits for all SPARC galaxies remain within 0.05 dex of the best-fit $\lc$ when the best-fit $\lm$ is held constant. Therefore, a truncated log-normal distribution with a 0.05 dex scatter, centered around the mean $\lc$ of the respective mock RC and truncated at 1$\sigma$, was considered for calculating the rotation velocity of individual data points within that mock RC.

It is worth noting that peculiar trends in the scatter are observed in the RCs of certain galaxies, where the data points consistently exhibit deviations below, above, or follow a sinusoidal pattern over several consecutive data points, as opposed to being randomly distributed around the mean fitted curve. These features may suggest the presence of non-circular motions or other systematic effects within the galactic disks \citep{Pineda_2016,Oman_2017}. However, addressing these effects falls outside the scope of this study. Our primary focus here is on the authenticity of the smooth DM component extracted from the RCs. The introduced random scatter in the mock RCs proves to be sufficient for capturing the overall complexities associated with fitting DM profiles to RCs.

\subsection{Generating mock RCs}

We created two types of mock RCs: (i) with only a DM component and (ii) with both DM and baryonic components. Below, we outline our general procedure for generating mock RCs.

For a mock RC of halo mass $M_{200}$, the concentration $c_{200}$ was derived from a truncated log-normal distribution centered around the mean MCR (eq.~\eqref{eq:s9}), with a scatter of 0.11 dex, and truncated at 2$\sigma$. This gave the DM component for the cuspy RCs. In the case of cored RCs, an additional parameter $r_c$ is needed, which was sampled as a random number between $0.3\,r_s$ and $0.9\,r_s$, where $r_s \equiv r_s(M_{200},c_{200})$ represents the scale radius for an NFW profile. This step generates the DM-only mock RCs.

\begin{figure}
    \centering
    \includegraphics[width = 0.7\linewidth]{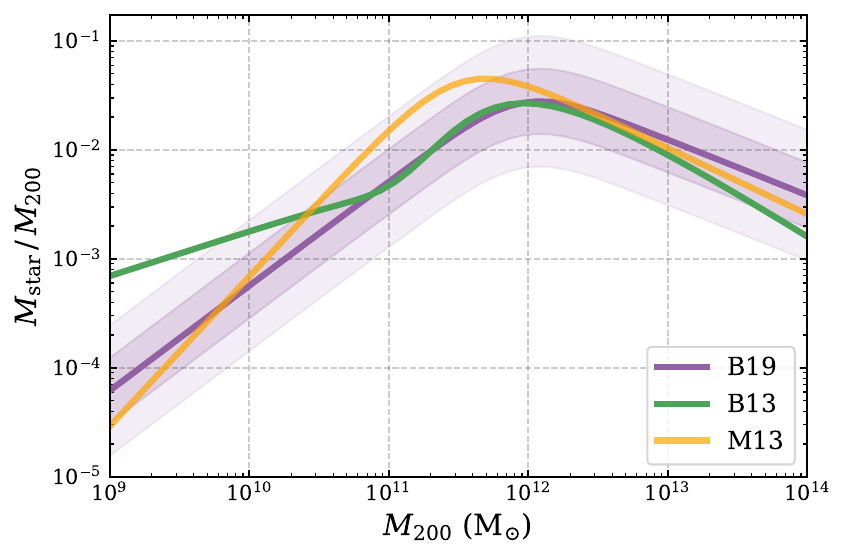}
    \caption{The ratio of stellar mass ($M_{\m{star}}$) to halo mass ($M_{200}$) as a function of $M_{200}$ for various stellar mass-halo mass (SMHM) relations used in this work. The purple line shows the SMHM relation from \citet{Behroozi_2019} (B19) along with its 1$\sigma$ and $2\sigma$ scatter as dark and light bands. The green line shows the SMHM relation from \citet{Behroozi_2013} (B13). For comparison, the orange line represents the SMHM relation from \citet{Moster_2013} (M13).}
    \label{fig:fig_2}
\end{figure}
The baryonic contributions were incorporated in the form of exponential disks for both the stellar and gas components, which require the total mass and disk scale of each component. The stellar mass, $\mstar$, for a given $M_{200}$, was obtained from one of the three SMHM relations illustrated in figure~\ref{fig:fig_2}: \citet{Behroozi_2019} (puple),\citet{Behroozi_2013} (green) and \citet{Moster_2013} (orange), hereafter referred to as B19, B13, and M13, respectively. The baryonic quantities, including $R_\m{star}$, $M_\m{gas}$, and $R_\m{gas}$, were acquired from the observed scaling relations as given in eq.~\eqref{eq:eq12}.

The radius ($r$) vs total circular velocity ($V_c$) data for a grid coordinate, $\grid$, are made according to $\Fi$, up to $\rsize \times r_s$. As mentioned, the scatter in $V_c$ is introduced by calculating $V_c$ at different $r$ by adding scatter to the $c_{200}$ within each mock RC. Error bars for each $V_c$ were derived from the distributions of the percentage error. By varying the mean $c_{200}$, scatter of $V_c$, and error bars of $V_c$, we created $N$ realizations of similar RCs for each of the two cuspy and cored DM profiles within a single grid of $\grid$. For the cored mock RCs, the ratio $r_c/r_s$ was uniformly sampled in the range (0.3, 0.9) for the $N$ realizations.

\subsubsection{Mock RCs with DM component only}
We simulate RCs assuming only the DM component is present. This assumption is far from reality, but it is important to understand the effect of radial extent, scatter, and error bars of RCs in distinguishing cores and cusps without the complications of baryonic components. Additionally, the results obtained from this simplified scenario serve as a control when examining subsequent simulations. For every $\grid$ pair, we generated a total of 20 realizations each for cuspy and cored mock RCs, resulting in a total of 9000 mock RCs.

\subsubsection{Mock RCs with DM and baryonic components}

We also generate mock RCs that include both DM and baryonic contributions. For a given $M_{200}$, $\mstar$ was calculated using B19 SMHM relation, along with its associated scatter. We use exponential disks for stellar and gas components, constraining them based on the observed scaling relations and the corresponding scatter as defined in eq.\eqref{eq:eq12}. For each $\grid$ pair, we generate $N$ = 50 realizations, resulting in a total of 22500 mock RCs.  

When considering a specific $M_{200}$, we introduced scatter to $M_{\m{star}}$ through a truncated log-normal distribution centered at the mean value specified by the B19 SMHM relation. As depicted in figure~\ref{fig:fig_2}, the B13 SMHM realtion exhibits a higher stellar-to-halo mass ratio in the lower mass range when compared to the B19 SMHM relation. To investigate the impact of the assumed SMHM relation on the final results, we conducted two additional sets of mock RC simulations of similar kinds but without the scatter in any baryonic parameters using both B19 and B13 SMHM relations and found that the final results remains the same. 

Additionally, $M_{\m{gas}}$, $R_{\m{gas}}$ and $R_{\m{star}}$ were sampled from a truncated log-normal distribution centered around the values from the relations in eq.~\eqref{eq:eq12}, including the observed scatter up to 2$\sigma$. The stellar velocity component, $\vstar$ for the mock RCs was scaled by a constant factor of $1/\sqrt{\yd}$, where $\m{log}(\yd)$ was sampled from a log-normal distribution around 0.5 with a scatter of 0.1 dex up to 3$\sigma$.

The scatter in baryonic scaling relation and SMHM relation incorporated results in diverse profiles at the inner regions of the mock RCs and can be used to directly verify the diversity problem observed in RCs, in the context of how variation in baryonic content can influence the RC diversity. We address this issue in a future work (\citet{MM_2024b}). In general, our mock RCs incorporate the diverse features of all the SPARC observed galaxies except for four outlier galaxies (F563-V1, F5742, IC2574, and UGC5750). The peculiarities in the cases of the two outliers, UGC5750 and IC2574 are also discussed in \citet{MM_2024c}.  

\subsection{Fitting mock RCs}
We fit all the simulated mock RCs with both cuspy and cored DM profiles using the $\Lambda$CDM set of priors. When fitting the mock RCs, we do not implement the regularization prior since obtaining a flat velocity for RCs not extending to the flat part is unrealistic. Otherwise, we follow the exact same fitting procedure conducted for the SPARC RCs. For DM-only mock RCs, only the DM component is fitted, while for mock RCs with baryonic components, $\yd$ is included as a free parameter. The best-fit value of a fit is recorded as the mean of its marginalized posterior, and its $1\sigma$ error is determined as the 68\% confidence band limits.

\section{Reliability of model selection and parameter estimation from RC fits}
\label{sec:rel}
In this section, we focus on the primary objective of this work, which is to assess the reliability of the DM density profiles inferred from model fitting to RCs. Our main goal is to identify regions in the $M_{200}$ - $\rsize$ parameter space where the cusp-core distinction, based on RC modeling, is most dependable. To initiate this analysis, we introduce the concept of the `reliability quotient', a metric we use to quantify the level of confidence in the model selection and parameter estimation process concerning RC fitting. Subsequent sections in this chapter explore the results of various sets of mock RCs, considering the reliability quotient and its dependence on different estimated halo parameters and observed galactic parameters.

Over the last two decades, extensive research has been dedicated to the challenge of deducing the internal characteristics of DM distribution in dwarf galaxies. Common techniques used to assess the ``coreness'' of a galaxy's central region include: (i) determining the inner slope of the radial density profile that best describes the central features in RCs \citep{Simon_2005,Oh_2015}, (ii) evaluating the inner mass deficit, which involves comparing the circular velocity at a fixed inner radius, such as $\vin (= V_c(2\,\mathrm{kpc}))$, to $\vmax$ or $\vlast (= V_c(\rlast))$ \citep{Naray_2008,Oman_2015}, and (iii) fitting the RCs by assuming various density profiles and employing model selection methods like reduced $\chisq$ comparisons \citep{deBlok_2008,Rodrigues_2017}. In some cases, it is possible to identify cores in galactic RCs through visual inspection of the rising segment, where a linear or gradual increase in RCs suggests a cored central region. Additionally, it is also possible to deduce and classify the presence of cores and cusps directly from the velocity fields from which the RCs are derived \citep{Naray_2011}.

The methods commonly employed to reveal the inner characteristics of DM profiles have proven valuable but have their own constraints. For instance, they might overlook complications arising from the presence of baryonic components, or the statistical power of model selection may not be strong enough to choose one profile over another. When comparing NFW density profiles to cored profiles in RCs, previous research indicated a statistical preference for the latter.  For instance, in studies by \citet{Katz_2017} and \citet{Ren_2019}, the distribution of reduced $\chi^2$ was used to show that cored profiles resulting from the SNe feedback hypothesis (DC14) and the SIDM hypothesis, respectively, were statistically favored over NFW profile. However, as \citet{Zentner_2022} pointed out, the traditional $\chi^2$ measure is not entirely accurate for this purpose \citep{Zentner_2022}. 

\citet{Zentner_2022} used Bayesian Information Criteria (BIC) for model selection between NFW, DC14, and SIDM density profiles in an equal setting and concluded that while there is a preference for cored profiles in low-mass galaxies, there is no systematic preference for either of the two cored models when considering galaxies of all masses. In such a scenario, where model selection on RCs fail to conclude the original mechanism behind cusp-core transformation, the most reliable methods for identifying the correct hypotheses involve: (i) examining the range of observed galaxies where DM cores are identified and comparing it with the range where a particular mechanism is efficient, and (ii) conducting a detailed study of the properties of the observed DM cores. To achieve this, one must confidently distinguish between cuspy and cored DM profiles from RCs and make robust measurements of the halo and core parameters.

Going beyond BIC, Bayes factor is a more robust model selection criteria that can break the degeneracy between cusp and core and between various cored profiles through mass modeling of RCs. Infact BIC can be shown to be an approximation of the Bayes factor \citep{Bhat_2010}. Expanding eq.~\ref{eq:eq11} using Laplace's method for approximating integrals, the evidence can be written as,
\begin{equation}
    \label{eq:bf_ex}
    \mathrm{ln}\mathcal{Z}_H(\bm{D}|H) = \mathrm{ln}\mathcal{L}(\Tilde{\bm{\uptheta}}|\bm{D}) + \mathrm{ln}\mathcal{G}(\Tilde{\bm{\uptheta}}) + \frac{|\bm{\uptheta}|}{2} \mathrm{ln}(2\pi) + \frac{1}{2}\mathrm{ln}|(-\mathcal{\mathcal{H}})^{-1}|,
\end{equation}
where $\mathcal{H}$ is the Hessian matrix and $\Tilde{\bm{\uptheta}}$ is the posterior mode where $\mathcal{L}(\bm{\uptheta}|H)\mathcal{G}(\bm{\uptheta})$ attains its maximum. If the number of data points, $n$, is large and assuming $\mathcal{G}(\bm{\uptheta}) = 1$, i.e, uninformative flat priors, the above expansion of ln$\mathcal{Z}_H$ reduces to,
\begin{equation}
    \label{eq:bf_bic}
    \mathrm{ln}\mathcal{Z}_H = \mathrm{ln}\mathcal{L}(\hat{\bm{\uptheta}}|\bm{D}) - \frac{|\bm{\uptheta}|}{2}\mathrm{ln}(n) = \mathrm{BIC},
\end{equation}
where $\Tilde{\bm{\uptheta}}$ is now replaced with the maximum likelihood estimator $\hat{\bm{\uptheta}}$ and the weak law of large numbers on the random variable $X_j = n \, \mathrm{ln}\mathcal{L}(\bm{\theta}|\bm{D})$ is invoked to fix $|-\mathcal{H}| = n^{|\bm{\uptheta}|}|\mathcal{I}|$, where $\mathcal{I}$ is the Fisher information matrix for a single data point. Note that the RC datasets commonly used like SPARC, having number of galaxies with low number of data points does not satisfy the criteria of large numbers. Neither are uniform priors often applicable, for e.g., when $\Lambda$CDM priors are used. Thus the criteria need for BIC to hold is not satisfied and one should use the full bayesian information contained.

\subsection{Reliability quotient}
As mentioned above, our primary goal is to identify whether there is a particular range of galaxies where cores are predominantly found, and whether this can be reliably claimed. To achieve this, we need to examine the collective behavior of an ensemble of mock RC realizations within specific parameter bins used for classifying these RCs. To start with, we consider the original $M_{200}$ and $\rsize$ as the parameters with which RCs are classified to study their reliability of cusp and cores. Then, for a given grid of $\grid$, we have an ensemble of $N$ mock RC realizations. 

One way of defining the reliability of cusp-core distinction, i.e, model selection, is to consider the arithmetic mean of the $\lbf$ values of all the mock RCs in this ensemble. However, this approach may be strongly influenced by the presence of extremely high or low $\lbf$ values, making it an inappropriate estimator of the ensemble's reliability. Alternatively, we can use the median, which provides a more robust estimate of the $\lbf$ for the ensemble. The median, while not influenced by extreme values, offers insights into only half of the mock RCs within an ensemble.

For cusp-core distinction and various other reliability quantification, we define a \textbf{reliability parameter} that determines the validity of the outcome of a given mock RC based on its fit results and its true nature, i.e, the original DM profile and the parameters assumed for making it. The reliability parameter, $\rela$, for the $i^\m{{th}}$ RC of an ensemble (i.e, fixed $\ori{M_{200}}$, $\ori{\rsize}$ and $\ori{\ipm}$) is defined as,
\begin{equation}
  \label{eq:r1}
  \rela(\ori{\ipm},\ori{M_{200}},\ori{\rsize}) = 
  \begin{cases}
    1, & \text{if } \ori{\ipm} = \text{cusp \& } \text{Condition-1} \\
    1, & \text{if } \ori{\ipm} = \text{core \& } \text{Condition-2} \\
    0, & \text{else}.
  \end{cases}
\end{equation}
Here $\ori{M_{200}}$ and $\ori{\rsize}$ represent the original values of $M_{200}$ and $\rsize$ with which the mock RC was made and $\ori{\ipm}$ represents the input model assumed for the DM density profile. In general, the conditions that have to be satisfied for the fit result of a mock RC to be reliable ($\rela = 1$) can be different for input cusp and input core mock RCs.

Using $\rela$ we define the \textbf{reliability quotient} for measuring the confidence in model selection and parameter estimation of RCs belonging to an ensemble (of a given model $\ipm$ and a particular range of $M_{200}$, $\rsize$) as,
\begin{equation}
  \label{eq:r2}
  \Rel(\ipm;M_{200},\rsize) =  \frac{1}{N}\sum_{i = 1}^{N}\rela(\ori{\ipm}=\ipm, \ori{M_{200}} = M_{200}, \ori{\rsize} = \rsize),
\end{equation}
where $N$ is the number of realizations of mock RCs in the ensemble with $M_{200}$ and $\rsize$ falling in the grid of $\origrid$ and with input model $\ori{\ipm}$. For a given ensemble, $\Rel\grid = 1$ implies that all the mock RCs belonging to it satisfied the reliability conditions used to define $\rela$. If we assume each realization in an ensemble is independent of each other\footnote{Since $P(A \cup B) = P(A) + P(B) - P(A \cap B)$ }, the reliability quotient $\Rel$ can be interpreted as the probability of success of model selection or parameter estimation, depending on the conditions of $\rela$, with a sufficiently large value of N. 

\subsection{Reliability of cusps and cores}
\label{ssec:Relcc}

For testing the authenticity of cores inferred from RCs, we make use of our various sets of mock RCs, which resemble the real data, by fitting them with both cuspy and cored DM profiles and employing rigorous model selection criteria. For this, we chose the powerful Bayes factor criteria for model selection. Unlike other Bayesian model selection methods that use quantities such as BIC, AIC, or $\chisq$, the Bayes factor ($\lbf$), defined in eq.~\eqref{eq:s10} and eq.~\eqref{eq:eq11}, takes into account the entire parameter space weighted by the likelihood and prior probabilities, implicitly accounting for the number of free parameters in the model and their correlations. With $\lbf$, the large number approximation on the number of data points required for the success of other statistical quantities could also be waived off, making it compatible with analyzing all the RCs of various numbers of data points.

\begin{figure*}
  \centering
  \includegraphics[width=1\linewidth]{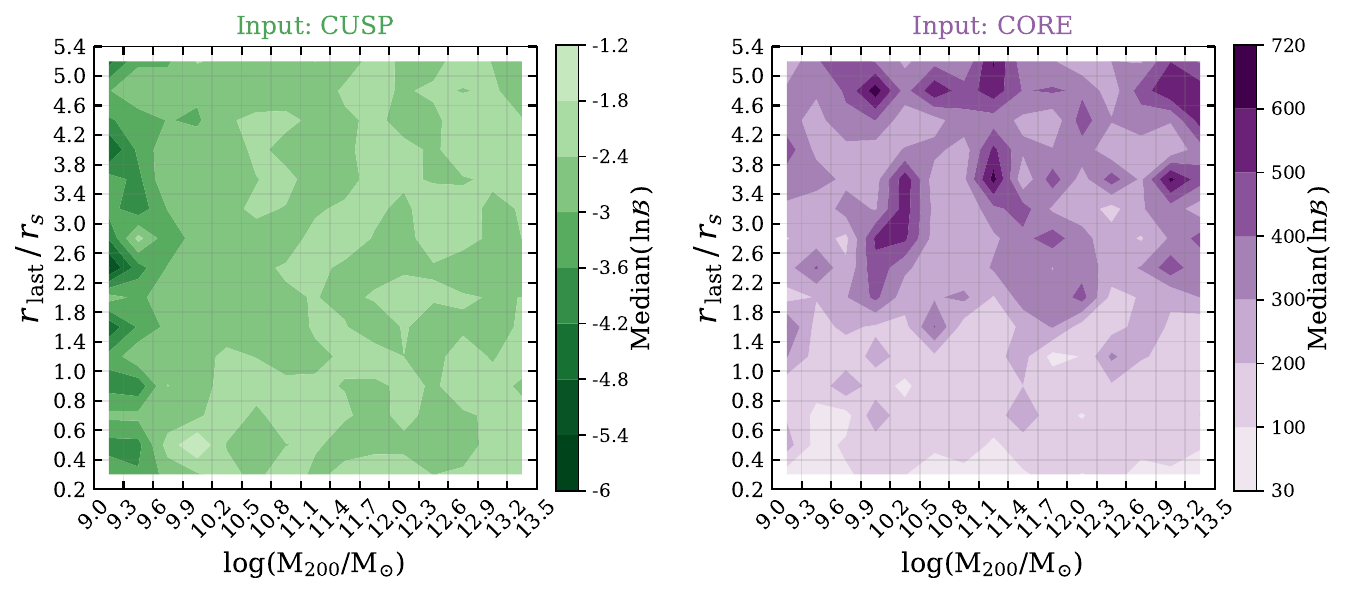}
  \caption{Median of the $\lbf$ values obtained from fitting cusp and core DM profiles to all the DM-only mock RCs of an ensemble (each grid). The left panel displays the results for input cusp mock RCs, while the right panel presents the results for input core mock RCs. Note the range of Median($\lbf$) in the left and right panels.}
  \label{fig:figr1}
\end{figure*}
Employing conditions on $\lbf$, we define the reliability parameter for cusp-core distinction as,
\begin{equation}
  \label{eq:r3}
  \rela(\lbf,\ori{\ipm},\ori{M_{200}},\ori{\rsize}) = 
  \begin{cases}
    1, & \text{if } \ori{\ipm} = \text{cusp \& }  \lbf \leq -1 \\
    1, & \text{if } \ori{\ipm} = \text{core \& }  \lbf \geq 5 \\
    0, & \text{else}.
  \end{cases}
\end{equation}
We apply different cutoffs on $\lbf$ for input cusp and input core mock RCs. This distinction arises from the fact that a cuspy RC can also be fitted by the cored DM density profile (eq.\eqref{eq:s6}) with a very small core size ($r_c$), thus necessitating stricter criteria to differentiate genuine cored RCs from cusp-like ones. This distinction becomes evident in the results of DM-only mock RCs, which do not involve the complexities introduced by baryonic components and are expected to effectively identify the true model. Figure~\ref{fig:figr1} illustrates the median values of $\lbf$ for each ensemble, consisting of 20 realizations for each $\ori{M_{200}}$ and $\ori{\rsize}$ pair, with input cusp (left panel) and input core (right panel) mock RCs. Note the significantly smaller value of $|\m{Median}(\lbf)|$ for input cusp mock RCs compared to input core mock RCs.


When $\rela = 1$, it indicates that the outcome of inferring the presence of a core in a specific mock RC is consistent with the original model assumed for its creation. Note that we do not question the reliability of the fitted parameters here. Rather, we are investigating the reliability of distinguishing between cusp and core DM profiles through model fitting to RCs. To quantify this reliability, we substitute the $\rela$ defined in eq.~\eqref{eq:r3} to eq.~\eqref{eq:r2} to calculate the reliability quotient for cusp-core distinction, denoted as $\Relcc$. This provides an estimate of the likelihood of a successful model selection between cuspy and cored DM profiles within a given ensemble of mock RCs.

\begin{figure*}
  \centering
    \includegraphics[width=1\linewidth]{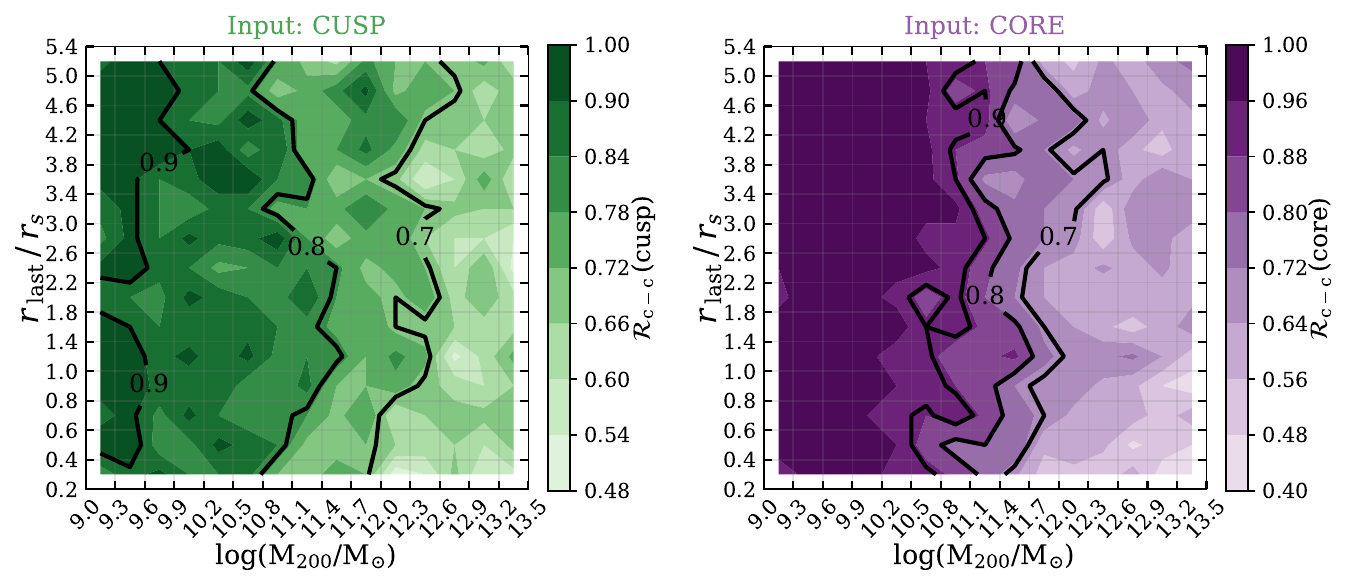}
  \caption{Reliability of cusp-core distinction ($\Relcc$) of RCs in an ensemble as a function of the grid parameters $\grid$. The left panel shows the results for input cusp mock RCs ($\ori{\ipm} = \m{cusp}$) and the right panel shows the result for input core mock RCs ($\ori{\ipm} = \m{core}$). The black contour lines enclose the 90\%, 80\%, and 70\% reliable regions.}
  \label{fig:figr2a}
\end{figure*}
In figure~\ref{fig:figr2a} we show the reliability of cusp-core distinction, $\Relcc$, which is determined for mock RCs featuring input cuspy profiles (in the left panel) and cored profiles (in the right panel). As previously explained, the challenge in distinguishing genuine cusps through model fitting to RCs arises from the potential of small-core cored profiles fitting cuspy RCs. This explains why, for a given mass range, $\Relcc(\ori{\ipm} = \mathrm{cusp})$ is less than $\Relcc(\ori{\ipm} = \mathrm{core})$. Nonetheless, for both cuspy and cored input mock RCs, $\Relcc$ decreases as $M_{200}$ increases. This necessitates the establishment of a cutoff on $M_{200}$, beyond which the reliability of model selection for cusps and cores becomes uncertain, depending on a lower bound on $\Relcc$.

\begin{figure*}
    \includegraphics[width=1\linewidth]{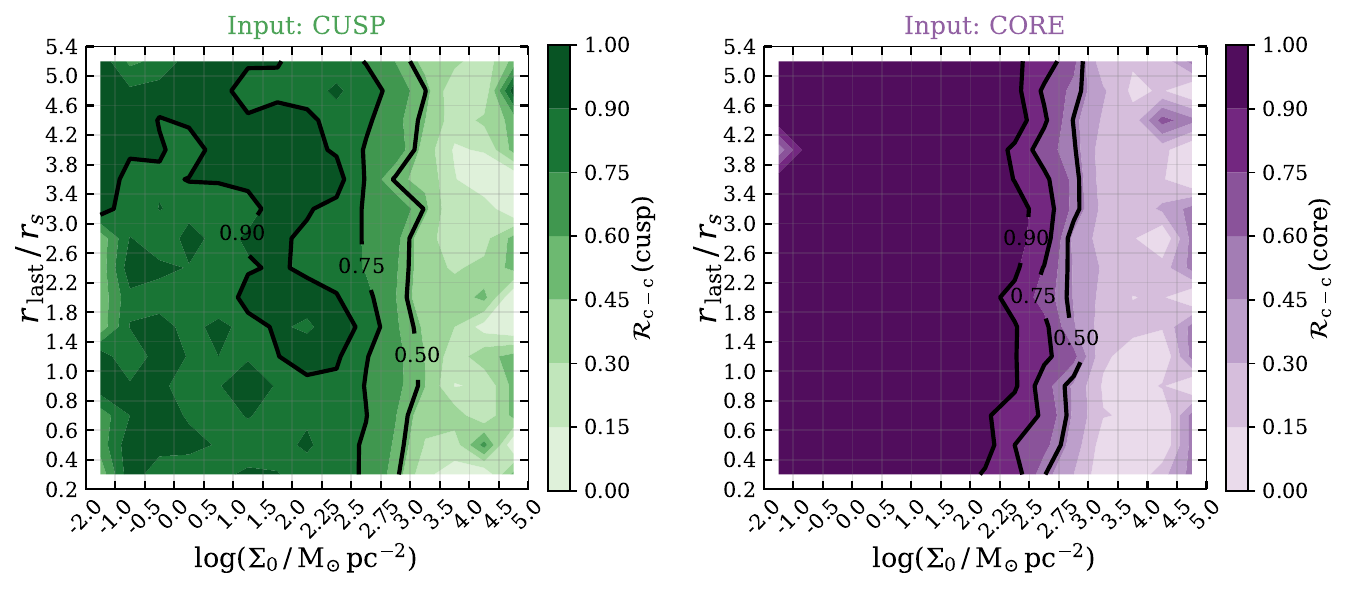}
  \caption{Reliability of cusp-core distinction ($\Relcc$) of RCs in an ensemble as a function of $(\Sigma_0,\rsize)$. The left panel shows the results for input cusp mock RCs and the right panel shows the result for input core mock RCs. The 90\%, 75\%, and 50\% reliable regions are marked in black contour lines.}
  \label{fig:figr2b}
\end{figure*}
In figure~\ref{fig:figr2b}, we present the reliability quotient as a function of $\Sigma_0$ instead of $M_{200}$.  This choice offers an explanation for the decline in $\Relcc$ concerning high-mass galaxies. High stellar surface brightness implies that the central regions of the RCs are primarily influenced by the stellar component, which can directly impact the fitting process for DM profiles due to the influence of $\yd$. Consequently, a truly cored RC could be mistakenly fitted with a cuspy profile and a slightly lower $\yd$, while conversely, a truly cuspy RC could be fitted with a cored profile and a slightly higher $\yd$.

\begin{figure*}
  \centering
  \includegraphics[width=1\linewidth]{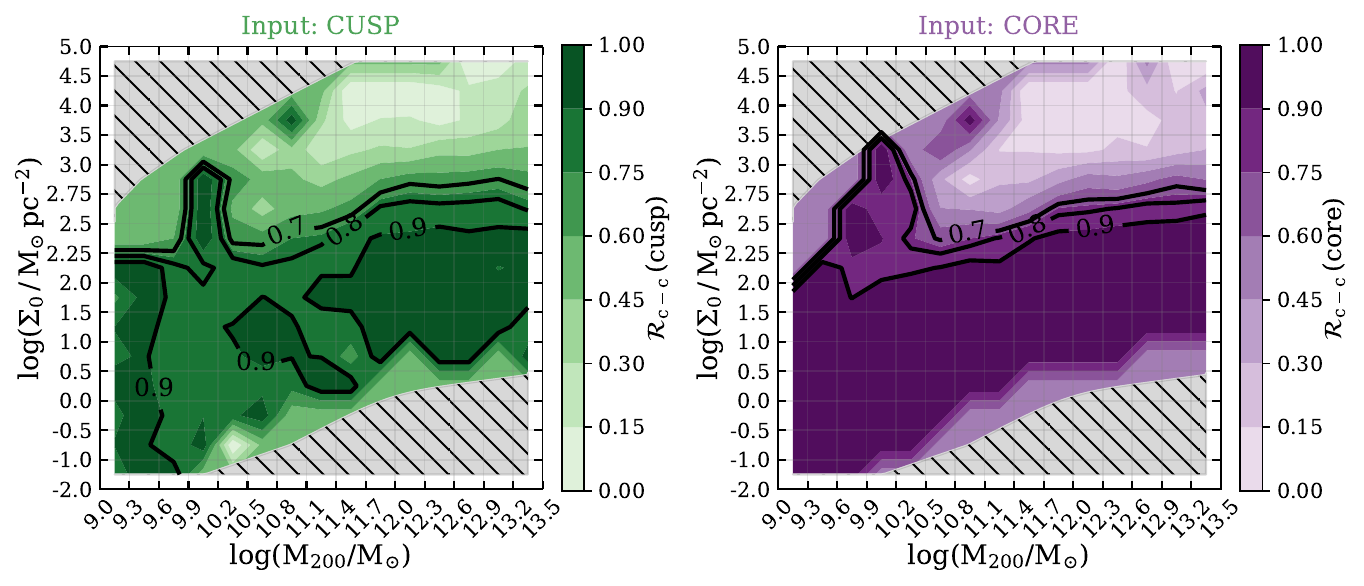}
  \caption{Reliability of cusp-core distinction ($\Relcc$) of RCs in an ensemble as a function of $(M_{200},\Sigma_0)$. The left panel shows the results for input cusp mock RCs and the right panel shows the result for input core mock RCs. The region shaded in grey are forbidden on account of the correlation between $\Sigma_0$ and $M_{200}$.}
  \label{fig:figr3}
\end{figure*}
As evident in figure~\ref{fig:figr2b}, the impact of the cutoff on the reliability of cusps and cores is more prominent with respect to $\Sigma_0$ compared to $M_{200}$. In the case of high $\Sigma_0$, $\Relcc$ decreases abruptly, as opposed to $\Relcc$ gradually decreasing with increasing $M_{200}$. In fact, $\Relcc \geq 0.5$ encompasses nearly the entire parameter space of $M_{200}$ while excluding the high $\Sigma_0$ region. For both $\Relcc(M_{200},\rsize)$ and $\Relcc(\Sigma_0,\rsize)$, there is no distinct cutoff on $\rsize$. In figure~\ref{fig:figr3}, $\Relcc$ is plotted as a function of $M_ {200}$ and $\Sigma_0$. Because $\Sigma_0$ is dependent on $M_{200}$, not all of the parameter space is available for real galaxies, and these forbidden regions are shaded in grey. figure~\ref{fig:figr3} shows that the cutoff under question depends only weakly on $M_{200}$ but strongly on $\Sigma_0$.

Since the true nature of an observed RC can be either cusp or core, the reliable region for cusp-core distinction should be obtained as the region where both input cuspy mock RCs and input cored mock RCs have a high $\Relcc$. We consider the reliable region to be set by the condition $\Relcc > 0.75$, which corresponds to the region where the true model of at least three-fourths of the mock RCs in an ensemble can be successfully recovered by model fitting. In terms of the rough probabilistic interpretation of $\Relcc$, this corresponds to the region where a cusp-core distinction could be made with a probability greater than 0.75, or the 75\% confidence level, with a 25\% chance of being incorrect. Since the dependence of these levels on $\rsize$ is weak, we can establish constant values for $\lms$ and $\lsigs$ as $\lms = 11.4 \pm 0.3$ and $\lsigs = 2.5 \pm 0.25$, based on $\Relcc(M_{200})$ and $\Relcc(\Sigma_0)$, which defines a threshold above which the reliability of cusp-core distinction, based on RC model fitting, cannot be assured.

Both these upper limits on $\lm$ and $\lsig$ depend on the required level of confidence set by the conditions on $\Relcc$. Lowering the $\Relcc$ cutoff increases the reliable $\lm$ space. Whereas, the upper limit on $\lsig$ more or less remains the same even if the condition on $\Relcc$ is relaxed from 0.75 down to 0.5. Consequently, $\Sigma_0$ is a critical factor in the reliability of cusps and cores inferred from RCs. This limitation needs consideration in studies that seek to compare the broad spectrum of galaxies where cores are observed and evaluate the effectiveness of cusp-core transformation mechanisms.  When inferring the presence of cusps or cores via model selection, it's crucial to recognize the potential for RCs from galaxies with genuine cores to be fitted with cuspy DM density profiles and vice versa. The upper limit on $\lsig$ of $2.5\pm0.25$ agrees with the cutoff mentioned in \citet{Zentner_2022}, where the preference for cores diminishes. This value also agrees with the boundary of two populations mentioned in \citet{Frosst_2022}, where the slow or steep rise of RC inner slopes are separated.

\subsection{Reliability of halo parameters}
\label{ssec:Relhp}
The reliability of the halo parameters ($M_{200}$ and $c_{200}$) derived from model fitting to mock RCs is considered in this section. We analyze two important necessities for the success of parameter estimation: the accuracy and precision of the fit results. To gauge the precision of a parameter $Q$, we introduce conditions based on $\pres{Q}$, where $Q$ and $\Delta Q$ denote the best-fit value and its associated 68\% error, respectively. These values are derived when an input cusp (core) mock RC is fitted while assuming a cuspy (cored) DM profile. On the other hand, accuracy is the measure of the deviation between the best-fit value and the original parameter value, obtained as $\accu{Q}$, with the bar signifying the true value used in the creation of the RC.

\begin{figure*}
  \centering
  \includegraphics[width = 1\linewidth, draft = false]{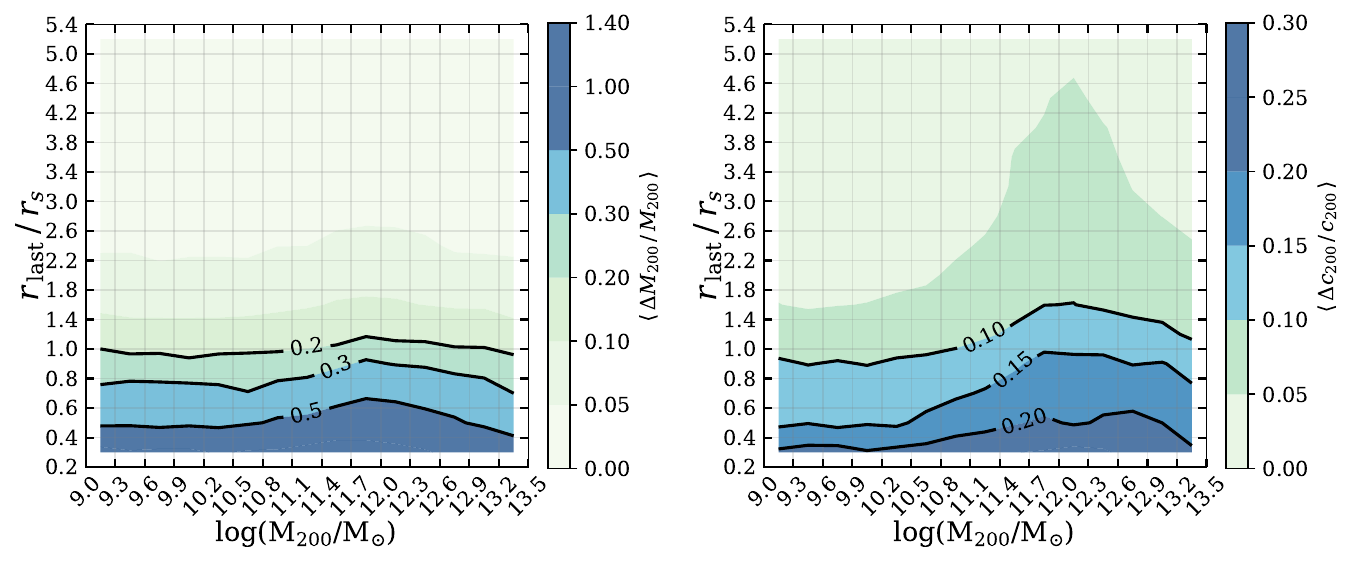}
  \caption{The arithmetic mean of $\pres{M_{200}}$ (in the left panel) and $\pres{c_{200}}$ (in the right panel) over an ensemble of mock RCs (each grid) obtained by fitting input cusp mock RCs with cuspy DM profile and input core mock RCs with cored DM profile.}
  \label{fig:figr4}
\end{figure*}
The left and right panels of figure~\ref{fig:figr4} display the arithmetic mean of $\pres{Q}$ for the halo parameters $M_{200}$ and $c_{200}$ as $\mean{\pres{M_{200}}}$ and $\mean{\pres{c_{200}}}$, respectively, taken over an ensemble of mock RCs. n this context, an ensemble encompasses all the mock RCs originating from both input cuspy and input cored models sharing the same $\origrid$.  It's evident from figure~\ref{fig:figr4} that $\rsize$ plays a decisive role in determining the precision of the fit results. As $\rsize$ increases, the radial extent of the RC span larger fractions of $r_s$, resulting in an increased number of data points in the mock RCs. Consequently, a smaller $\rsize$ leads to greater uncertainties in the halo parameters, as reflected in the trend of increasing $\mean{\pres{Q}}$ with decreasing $\rsize$. 

This trend is particularly evident for $M_{200},$ which is predominantly influenced by the flat part of the RCs. However, when considering $c_{200},$ which is sensitive to the inner characteristics of the RCs, the influence of baryons becomes more pronounced. This explains the rise in $\pres{Q}$ as we move toward higher-mass regions, where the impact of baryonic components on the inner sections of the RCs becomes more substantial. Importantly, the observed trends in $\pres{M_{200}}$ and $\pres{c_{200}}$ in figure~\ref{fig:figr4} hold true whether we consider input cuspy and input cored mock RCs separately.

\begin{figure*}
  \centering
  \includegraphics[width = 1\linewidth, draft = false]{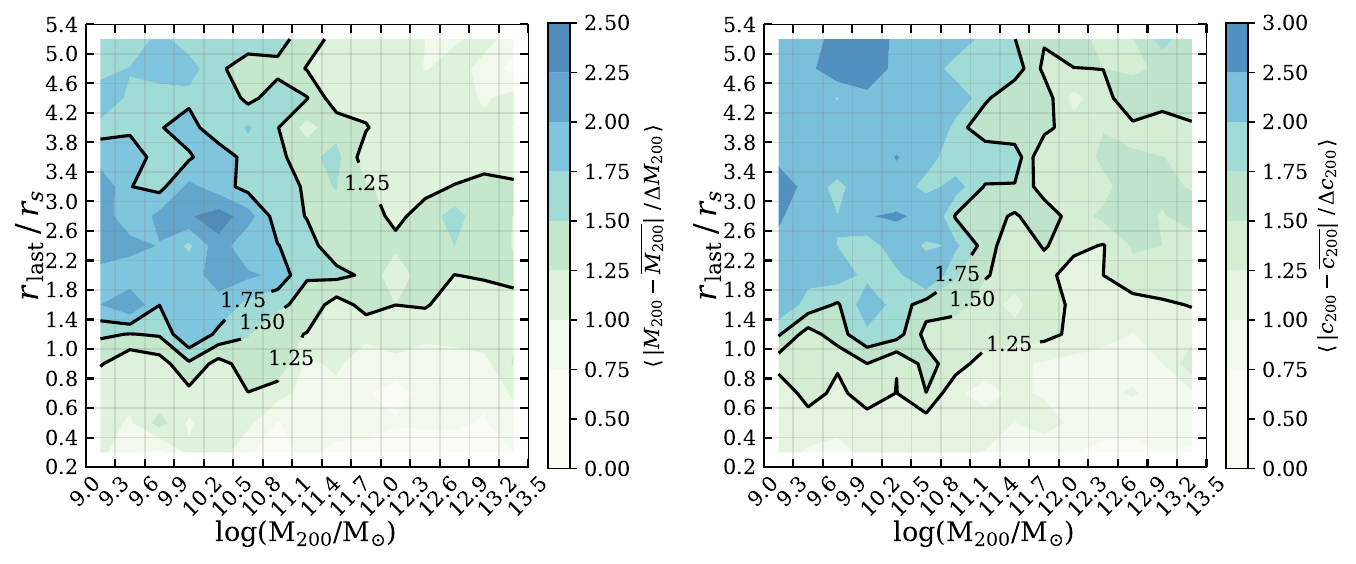}
  \caption{The arithmetic mean of $\accu{M_{200}}$ (in the left panel) and $\accu{c_{200}}$ (in the right panel) over an ensemble of mock RCs (each grid) obtained by fitting input cusp mock RCs with cuspy DM profile and input core mock RCs with cored DM profile.}
  \label{fig:figr5}
\end{figure*}
In figure~\ref{fig:figr5}, we illustrate the arithmetic mean of $\accu{Q}$ over an ensemble, denoted as $\mean{\accu{Q}},$ for both the halo parameters. Once again, this ensemble encompasses all mock RCs with the same $\origrid$ originating from both input cuspy and cored models. These trends persist whether we analyze both input models separately or collectively.  The accuracy, in this case, represents the number of $\sigma$'s the original parameter is away from the best fit value of the parameter, where $1\sigma$ is the 68\% confidence band given by $\Delta Q$. 

The left panel of figure~\ref{fig:figr5} outlines the contours of $\mean{\accu{M_{200}}}$ (black lines) denoting the regions where the best-fit values of $M_{200}$ are, on average, within 1.25$\sigma$, 1.5$\sigma$, and 1.75$\sigma$ of $\ori{M_{200}}$. Similarly, these contours are presented for $c_{200}$ in the right panel. These contours show that the accuracy is high for low $\rsize$ RCs, which is to be expected given that, as previously discussed, $\Delta Q$ increases with lower $\rsize$. This also explains why the accurate $c_{200}$ regions are on the right side of the plot, where the $\Delta c_{200}$ is large. 

In the case of RCs with low $M_{200}$, both $\mean{\accu{M_{200}}}$ and $\mean{\accu{c_{200}}}$ exhibit high values.  A mixture of the scatter and the absolute error in the data determines the final accuracy with which the underlying parameters can be obtained. In RCs with low $M_{200}$, where the DM component dominates, scatter in data points with small $\Delta V_c$ values play a pivotal role in determining the ultimate fit results. Consequently, the scatter in such data points exerts a significant impact on the overall outcome.

We define the reliability parameter, $\rela$ to assess the reliability of halo parameter estimation from RC fits using conditions on $\pres{Q}$ and $\accu{Q}$,
\begin{equation}
  \label{eq:r4}
  \rela(\ori{M_{200}},\ori{\rsize}) = 
  \begin{cases}
    1, & \text{if, } \pres{Q} \leq \m{Precision}(Q) \text{ \& }
        \accu{Q} \leq \m{Accuracy}(Q) \\
       & \,\,\,\,\,\, \text{ for } Q \in \{M_{200},c_{200}\}\\
    0, & \text{else.}
  \end{cases}
\end{equation}
Here, $Q$ represents the best fit value, and $\Delta Q$ represents its 68\% confidence band, which is determined by fitting the original model used to create the mock RC. The ensemble defined by $\origrid$ includes both cuspy and cored mock RCs, and the same conditions are applied for $\ori{\ipm} = $ cusp and $\ori{\ipm} = $ core.

\begin{figure*}
  \centering
  \includegraphics[width=1\linewidth, draft = false]{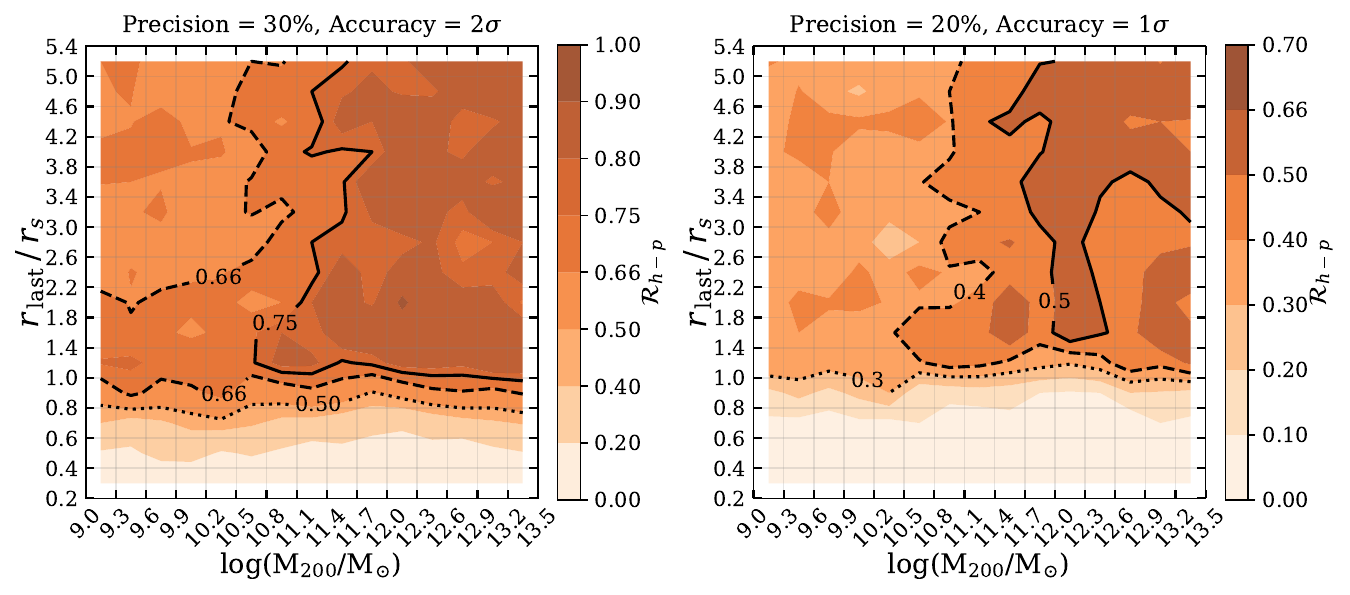}
  \caption{Reliability of halo parameters estimation ($\Relhp$) of RCs in an ensemble as a function of the grid parameters $\grid$. The left panel shows the results with a precision cutoff of 30\% and an accuracy cutoff of 2$\sigma$. The right panel shows the result with a precision cutoff of 20\% and an accuracy cutoff of 1$\sigma$. The lines enclose various reliable region as noted in the contours.}
  \label{fig:figr6}
\end{figure*}
By substituting $\rela$ from eq.~\eqref{eq:r4} into eq.~\eqref{eq:r2} and specifying precision and accuracy cutoffs, we derive the reliability quotient, $\Relhp$, which is employed to evaluate the success of halo parameter estimation using RC fits. Figure~\ref{fig:figr6} displays the plot of $\Relhp\grid$ for two different cases of precision and accuracy conditions on the halo parametersThe left panel shows $\Relhp$ obtained with lenient requirements of $\pres{Q} \leq 0.3$ and $\accu{Q} \leq 2$, while the right panel illustrates $\Relhp$ obtained with stricter criteria of $\pres{Q} \leq 0.2$ and $\accu{Q} \leq 1$. In both cases mock RCs with lower $\rsize$ are unreliable, with the specific lower limit on $\rsize$ for the reliable regions depending on the chosen cutoff for $\Relhp$ and the required precision in the halo parameter measurements.

The requirement for high precision and accuracy leads to reduced reliability in the estimation of fitted halo parameters, as evident from the generally lower values of $\Relhp$ in the right-hand panel compared to the left-hand panel of figure~\ref{fig:figr6}. While precision primarily imposes a lower limit on $\rsize$ below which the reliability of the estimated halo parameters diminishes, the requirement for accuracy impacts $\Relhp$ across the entire parameter space of $\grid$. In general, the high $\rsize$ region below $\lms = 11.1$ has lower accuracy and should be considered unreliable. The reliable region for halo parameters complements the region where cusp-core distinctions can be made reliably, indicating that small-scale effects like cusps and cores have limited influence on the inference of large-scale halo characteristics.

\begin{figure*}
  \centering
  \includegraphics[width=1\linewidth, draft = false]{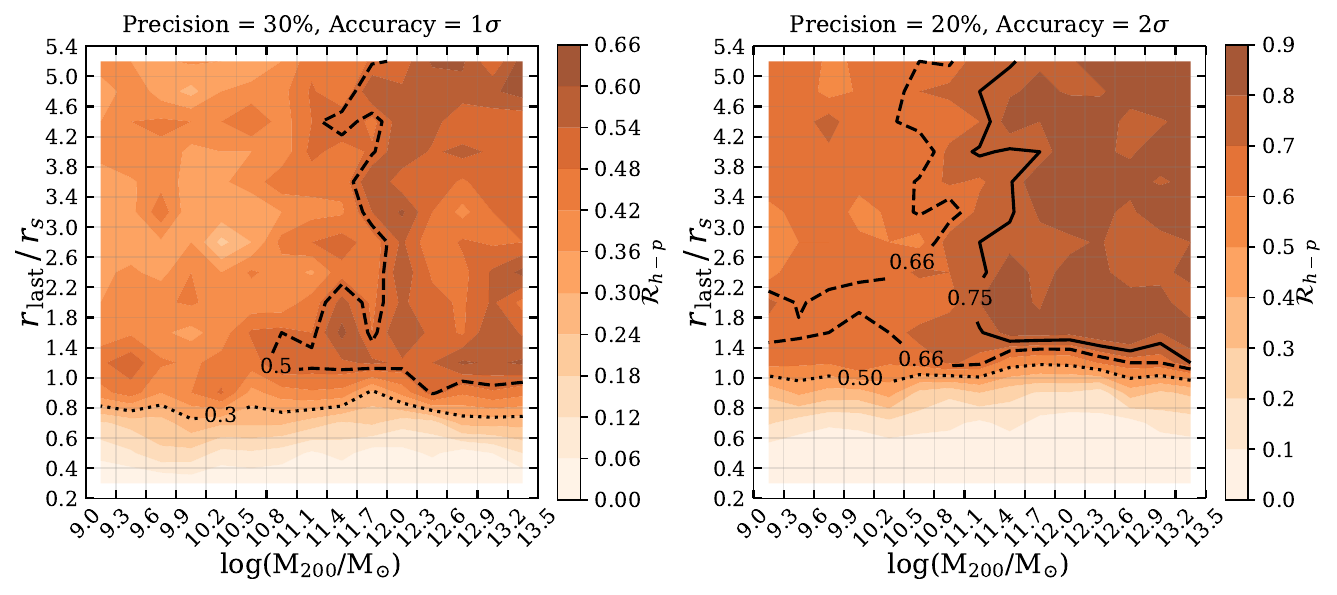}
  \caption{Same as figure~\ref{fig:figr6} but with the accuracy cutoff interchanged. The left panel shows $\Relhp$ with the precision cutoff of 30\% and accuracy cutoff of 1$\sigma$. The right panel shows the result with a precision cutoff of 20\% and an accuracy cutoff of 2$\sigma$.}
  \label{fig:figr7}
\end{figure*}
The plots in figure~\ref{fig:figr7}, which correspond to those in figure~\ref{fig:figr6} but with interchanged accuracy cutoffs, illustrate the impact of accuracy and precision on the reliability quotient. In the left panel of figure~\ref{fig:figr7}, $\Relhp$ is depicted for low precision (30\%) and high accuracy (1$\sigma$), while the right panel shows $\Relhp$ for higher precision (20\%) and low accuracy (2$\sigma$). Strengthening the accuracy criteria results in a decrease in the overall reliability, as observed when comparing the left panels of Figures~\ref{fig:figr6} and \ref{fig:figr7}. Conversely, the less reliable regions above $\rsize = 1$, obtained under high precision and high accuracy conditions in the right-hand panel of figure~\ref{fig:figr6}, become more reliable as accuracy requirements are relaxed from 1$\sigma$ to 2$\sigma$.

In conclusion, the reliable regions within $\grid$ for estimating halo parameters through model fitting to RCs are determined by a lower limit on $\rsize$ across the entire $\lm$ space and an upper limit on $\rsize$ for cases where $\lms \lesssim 11.1$. The exact value of the lower and upper limits on $\rsize$ depends on the precision and accuracy one demands for the fitted halo parameters and the required confidence in the reliability of the estimation process. However, it is evident that the most reliable regions for halo parameter estimation via RC fitting are found in high-mass galaxies with large $\rsize$. In contrast, for low-mass galaxies, RCs of the current quality are insufficient to obtain accurate measurements of their halo parameters. To properly model the halo parameters of low-mass galaxies, RCs with higher quality, characterized by lower error bars, reduced scatter, and better resolution than the average properties of the SPARC RCs, are imperative.

\subsection{Reliability of core parameters}

As previously mentioned, comprehending the characteristics of central cores within DM density profiles of disk galaxies is crucial for identifying the mechanisms responsible for cusp-core transformation. Two key core parameters that can be inferred from RCs are the central core density, $\rho_b$, and the core size, $r_c$. While the question of the reliability of these core parameters, as determined through parameter estimation on RC data, is indeed contingent on the prerequisites for the successful distinction between cusps and cores, in this section, the reliability of core parameter estimation ($\rho_b$ and $r_c$) from RC fits will be explored assuming we could correctly identify a given RC that has cored DM density profile, and as such, only mock RCs with input-cored models will be considered.

\begin{figure*}
  \centering
  \includegraphics[width=1\linewidth, draft = false]{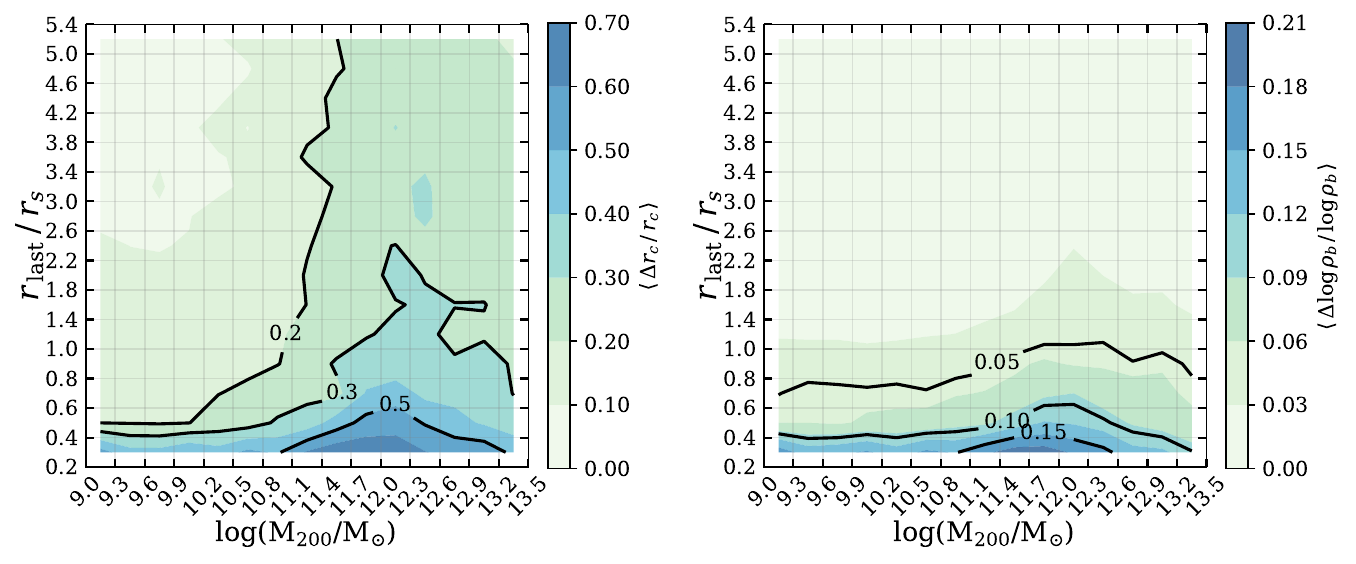}
  \caption{The arithmetic mean of $\pres{r_c}$ (in the left panel) and $\pres{\lrhob}$ (in the right panel) over an ensemble of mock RCs (each grid). Only the results of cored mock RCs fitted with the cored profile are considered here.}
  \label{fig:figr8}
\end{figure*}
\begin{figure*}
  \centering
  \includegraphics[width=1\linewidth, draft = false]{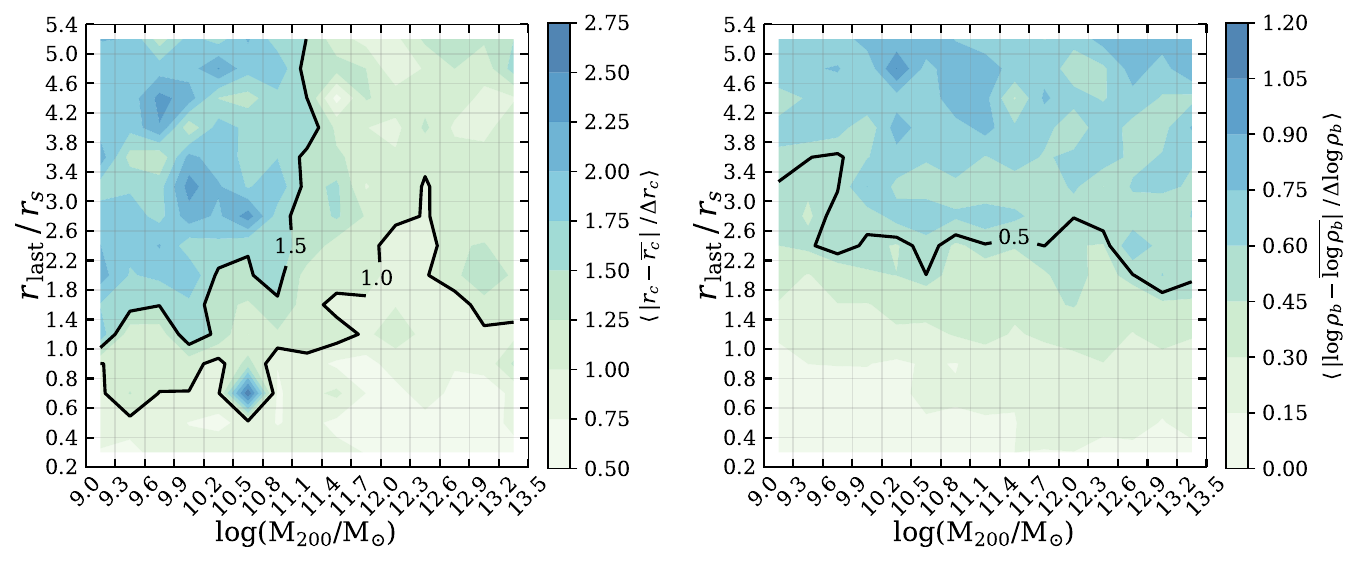}
  \caption{The arithmetic mean of $\accu{r_c}$ (in the left panel) and $\accu{\lrhob}$ (in the right panel) over an ensemble of mock RCs (each grid). Only the results of cored mock RCs fitted with cored profile are considered here. }
  \label{fig:figr9}
\end{figure*}
Following the approach used for halo parameters, we apply criteria to the precision and accuracy of the estimated core parameters obtained through fitting a cored profile to input cored mock RCs. Figures~\ref{fig:figr8} and \ref{fig:figr9} present the values of $\mean{\pres{Q}}$ and $\mean{\accu{Q}}$, respectively, where the left panels display the results for $Q = r_c$, and the right panels present the results for $Q = \lrhob$. Similar to the trends observed for the halo parameters, the precision of the fit results for both $r_c$ and $\lrhob$ decreases for low $\rsize$ RCs. Furthermore, for $r_c$, precision decreases for high $\lm$ RCs, as expected, since the baryonic components in the inner regions of such RCs influence the estimation of $r_c$.

The accuracy of $r_c$, as measured by $\accu{r_c}$ and depicted in the left panel of figure~\ref{fig:figr9}, is higher for low $\rsize$ and high $\lm$ RCs. This is exactly the region in which $r_c$ precision is low, implying that the high accuracy is due to a large $\Delta r_c$ rather than $r_c$ being close to $\ori{r_c}$. A similar trend is seen for the accuracy of $\lrhob$, right panel of figure~\ref{fig:figr9}, for which the regions of high accuracy coincide with the low precision regions. We use the reliability parameter defined in eq.\ref{eq:r4} with $Q \in {r_c, \lrhob}$, and substitute it into eq.\ref{eq:r2} to obtain the reliability quotient $\Relcp$, to quantify the reliability of the estimated core parameters.

\begin{figure*}
  \centering
  \includegraphics[width=1\linewidth, draft = false]{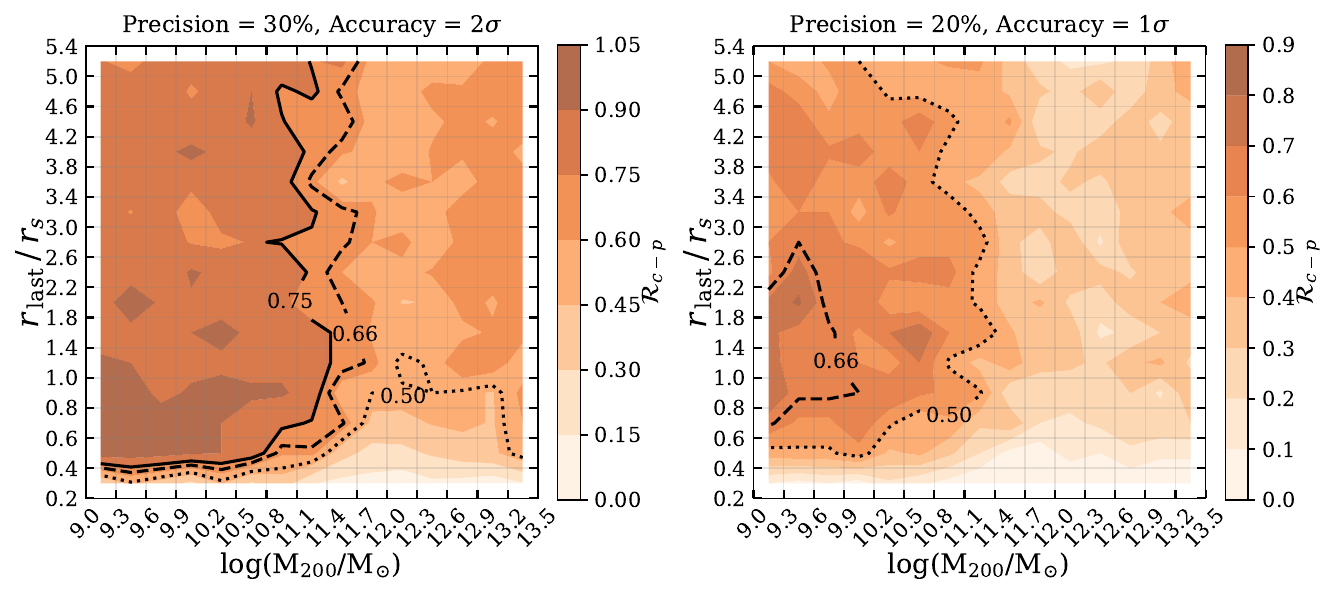}
  \caption{Reliability of core parameters estimation ($\Relcp$) of RCs of in an ensemble as a function of the grid parameters $\grid$. The left panel shows the results with the precision cutoff of 30\% and accuracy cutoff of 2$\sigma$. The right panel shows the result with a precision cutoff of 20\% and an accuracy cutoff of 1$\sigma$. The black lines enclose various reliable regions as noted in each contour.}
  \label{fig:figr10}
\end{figure*}
The resulting plot of $\Relcp$ with two sets of conditions on the precision and accuracy of the fitted core parameters is displayed in figure~\ref{fig:figr10}. In the left panel, $\Relcp$ is presented with lenient precision and accuracy criteria of $\pres{Q} \leq 0.3$ and $\accu{Q} \leq 2$. The right panel shows the resulting $\Relcp$ with stricter criteria of $\pres{Q} \leq 0.2$ and $\accu{Q} \leq 1$. These plots highlight that the most reliable range of RCs for estimating core parameters is characterized by low $\lm$ and high $\rsize$ RCs. The upper bound on $\lm$ aligns with the upper limit identified for the reliability of the cusp-core distinction, while the lower bound on $\rsize$ depends on the specific precision cutoff used to define $\Relcp$.

\section{Discussion}
\label{sec:gal}

\subsection{Reliable SPARC RCs}
\begin{figure*}
    \centering
    \includegraphics[width = 1\linewidth]{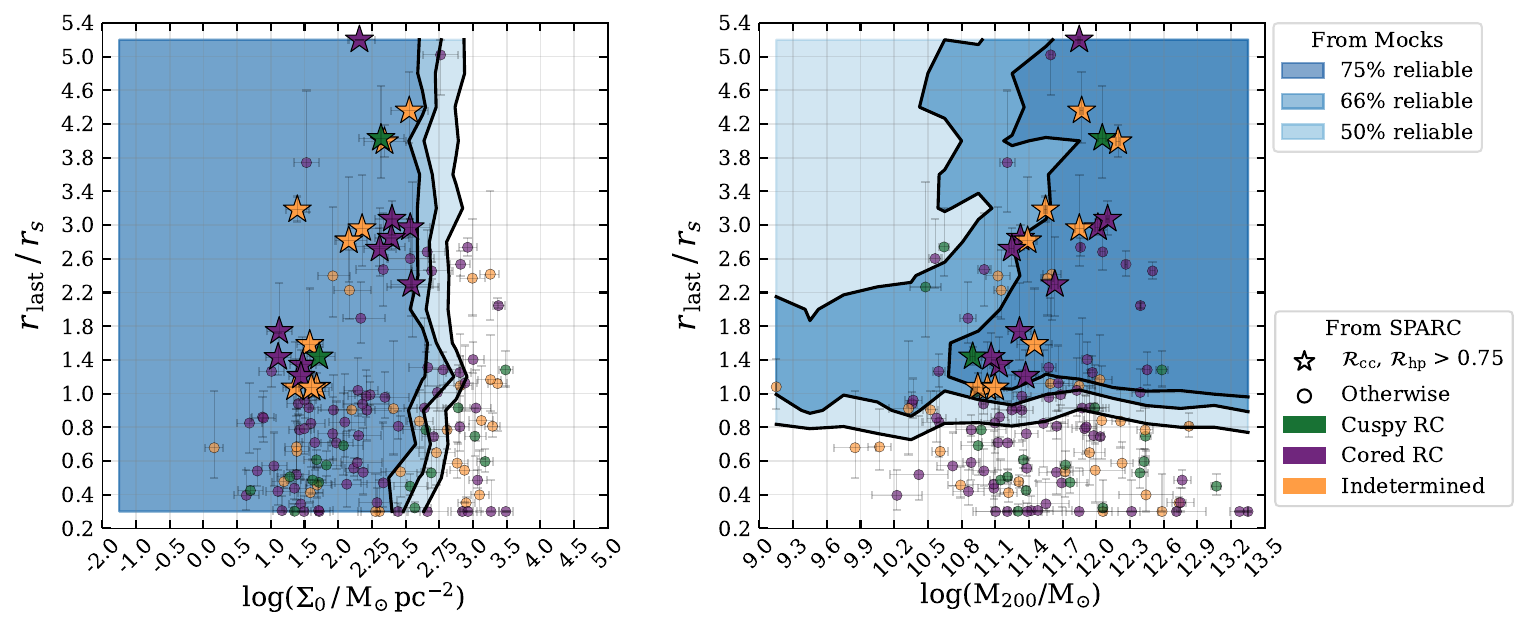}
    \caption{The above plot shows where the real observed SPARC galaxies fall in the `reliability-space'. The 75\% (dark blue contours), 66\% (medium blue contours), and 50\% (light blue contour) are the reliable regions, obtained from the mocks, for either cusp-core distinction as a function of $(\lsig,\rsize)$ [left panel], or the halo parameters estimation as a function of $\grid$ [right panel]. On top of these, the SPARC galaxies are shown with scattered symbols. Stars (circles) represent SPARC galaxies with $> 75$\% ($< 75$\%) reliability for {\it both} cusp-core and halo-parameters together. The colors of the markers are used to represent the best underlying DM halo inferred from the RCs based on the value of the Bayes factor - green shows cuspy DM profiles, purple shows cored DM profiles, and orange represents DM profiles that cannot be confidently assigned to being either cuspy or core.\\ Note, that in either panel, there are points with (say) $<50$\% reliability that falls in regions with a higher reliability since for observed galaxies we are looking at the reliability in both cusp-core and halo parameters. Plots for reliable SPARC galaxies for only cusp-core or halo parameter are shown in appendix~\ref{sec:appB}.
    }
    \label{fig:fig_15}
\end{figure*}
In this section, we apply the findings from the previous section to the observed SPARC RCs. Figure~\ref{fig:fig_15} provides a summary, with SPARC RC positions plotted alongside reliability graphs. In the left panel, the reliability plot for distinguishing cusps and cores by fitting cuspy and cored DM density profiles to RC data is displayed on the $(\m{log}\Sigma_0,\rsize)$ plane. The 75\%, 66\%, and 50\% reliable regions of cusp-core distinction are shaded with darker to lighter blue bands, respectively.

It was noted in Section~\ref{ssec:Relcc} that the reliable region for cusp-core distinction was independent of $\rsize$. We point out that, this independence on $\rsize$ could be, at least partly, the effect of the $\Lambda$CDM prior implemented in the fitting procedure for the mock RCs, since this disregards fitting core RCs of small extensions with wild cuspy profiles mimicking such inner RC characteristics. Therefore, here we consider SPARC RC fit results obtained by implementing the $\Lambda$CDM prior along with the regularization prior of $\vmax$. 

There are 89 SPARC RCs within the $\Relcc > 0.75$ region, 10 RCs within $0.75 > \Relcc > 0.66$, and 9 RCs with $0.66 > \Relcc > 0.5$. In figure~\ref{fig:fig_15}, green points denote RC fits with $\lbf < -1$, representing cuspy RCs. Purple points correspond to fits with $\lbf > 1$, representing cored RCs, while orange points represent RCs with $-1 < \lbf < 1$, where a clear distinction of cusp or core is not . In the 75\% reliable region, there are 13 clearly cuspy RCs, 54 clearly cored RCs, and 22 RCs for which their true model remains undetermined, given that these results are inherently sensitive to our fitting process and model selection criteria.

The reliability of halo parameters, as shown in the right panel of figure~\ref{fig:fig_15}, is marked using the same color scheme as in the left panel to indicate the various reliable regions on the $\grid$ plane. In the SPARC dataset, there are 36 galaxies whose RCs fall within the region with $\Relhp > 0.75$, 22 galaxies with RCs in the $0.75 > \Relhp > 0.66$ region, and 14 galaxies with RCs in the $0.66 > \Relhp > 0.50$ region. These reliability values were obtained with the criteria that precision is 30\% or less and accuracy is within 2$\sigma$.

We have identified 21 galaxies in the SPARC dataset that fall into regions of both cusp-core distinction and halo parameter estimation with reliability values greater than 0.75. These galaxies are marked with stars in figure~\ref{fig:fig_15}, signifying that the model selection and halo parameter estimation results are highly reliable for them. Additionally, there are 23 galaxies that fall within the region where $0.75 > \Relcc$ and $\Relhp > 0.66$ and 15 galaxies have RCs falling in the $0.66 > \Relcc$ and $\Relhp < 0.5$ region (marked with circular points in figure~\ref{fig:fig_15}. Note that around 55\% of the SPARC RCs fall outside of the 50\% reliable region in either of the cusp-core distinction or halo-parameter estimation plots. 

\begin{table*}
\small
\centering
\caption{Fit results of the most reliable sample of RCs from SPARC, which falls on the region where $\Relcc > 0.75$ and $\Relhp > 0.75$.}
\label{tab:table_2}
\begin{adjustbox}{max width=1.1\textwidth,center}
\begin{tabular}{|lllllllllr|}
\hline
 Galaxy   & Model   & $\lbf$   & $\lms$           & $c_{200}$        & $r_c$            & $\yd$           & $\lmstar$        & $\m{log}(\Sigma_0)$   &   $\rsize$ \\
\hline
\multicolumn{10}{| c |}{Cuspy RCs}\\
\hline
 UGC12506 & cusp    & -1.57    & $12.05 \pm 0.04$ & $17.24 \pm 1.90$ & -                & $0.54 \pm 0.10$ & $10.85 \pm 0.08$ & 2.3                   &       4.03 \\
  & core    &          & $12.05 \pm 0.04$ & $17.71 \pm 1.93$ & $2.10 \pm 1.62$  & $0.53 \pm 0.10$ & $10.84 \pm 0.08$ &                       &       4.16 \\
 UGC02259 & cusp    & -1.15    & $10.90 \pm 0.02$ & $15.42 \pm 0.90$ & -                & $0.52 \pm 0.12$ & $8.93 \pm 0.10$  & 1.7                   &       1.44 \\
  & core    &          & $10.90 \pm 0.02$ & $15.56 \pm 0.89$ & $0.92 \pm 0.53$  & $0.51 \pm 0.12$ & $8.92 \pm 0.10$  &                       &       1.45 \\
 \hline
 \multicolumn{10}{| c |}{Model undetermined RCs}\\
 \hline
 F568-V1  & cusp    & -0.42    & $11.45 \pm 0.16$ & $12.59 \pm 2.14$ & -                & $0.54 \pm 0.12$ & $9.29 \pm 0.10$  & 1.6                   &       1.59 \\
   & core    &          & $11.38 \pm 0.15$ & $13.93 \pm 2.32$ & $3.62 \pm 1.97$  & $0.54 \pm 0.13$ & $9.29 \pm 0.10$  &                       &       1.86 \\
 NGC0289  & cusp    & -0.11    & $11.85 \pm 0.04$ & $7.69 \pm 1.12$  & -                & $0.58 \pm 0.06$ & $10.63 \pm 0.05$ & 2.2                   &       2.96 \\
   & core    &          & $11.85 \pm 0.04$ & $7.68 \pm 1.11$  & $9.72 \pm 5.64$  & $0.59 \pm 0.06$ & $10.64 \pm 0.05$ &                       &       2.96 \\
 UGC00128 & cusp    & -0.08    & $11.55 \pm 0.01$ & $8.48 \pm 0.34$  & -                & $0.47 \pm 0.09$ & $9.74 \pm 0.09$  & 1.4                   &       3.18 \\
  & core    &          & $11.55 \pm 0.01$ & $8.54 \pm 0.34$  & $5.37 \pm 2.65$  & $0.46 \pm 0.09$ & $9.73 \pm 0.09$  &                       &       3.21 \\
 NGC5985  & cusp    & 0.43     & $12.19 \pm 0.02$ & $27.04 \pm 1.10$ & -                & $0.31 \pm 0.05$ & $10.84 \pm 0.06$ &                       &       4.01 \\
   & core    &          & $12.19 \pm 0.02$ & $27.02 \pm 1.08$ & $7.21 \pm 1.52$  & $0.29 \pm 0.05$ & $10.83 \pm 0.07$ & 2.3                   &       3.99 \\
 NGC5033  & cusp    & 0.74     & $11.86 \pm 0.02$ & $19.57 \pm 1.50$ & -                & $0.38 \pm 0.05$ & $10.70 \pm 0.05$ &                       &       4.77 \\
   & core    &          & $11.87 \pm 0.02$ & $18.02 \pm 1.54$ & $8.10 \pm 2.22$  & $0.42 \pm 0.06$ & $10.75 \pm 0.05$ & 2.5                   &       4.36 \\
 UGC07399 & cusp    & 0.82     & $11.22 \pm 0.11$ & $16.35 \pm 1.82$ & -                & $0.53 \pm 0.12$ & $8.81 \pm 0.10$  &                       &       0.88 \\
  & core    &          & $11.10 \pm 0.11$ & $18.73 \pm 2.28$ & $1.65 \pm 0.50$  & $0.56 \pm 0.13$ & $8.83 \pm 0.10$  & 1.6                   &       1.10 \\
 F571-V1  & cusp    & 0.84     & $11.04 \pm 0.17$ & $8.02 \pm 1.87$  & -                & $0.49 \pm 0.11$ & $8.94 \pm 0.10$  &                       &       1.03 \\
   & core    &          & $11.03 \pm 0.15$ & $8.31 \pm 1.63$  & $7.31 \pm 3.49$  & $0.50 \pm 0.11$ & $8.94 \pm 0.10$  & 1.4                   &       1.10 \\
 NGC3769  & cusp    & 0.88     & $11.40 \pm 0.08$ & $9.94 \pm 1.63$  & -                & $0.39 \pm 0.06$ & $9.87 \pm 0.07$  &                       &       2.79 \\
   & core    &          & $11.39 \pm 0.08$ & $9.90 \pm 1.50$  & $8.23 \pm 3.38$  & $0.42 \pm 0.06$ & $9.91 \pm 0.07$  & 2.1                   &       2.82 \\
 UGC04325 & cusp    & 0.99     & $10.98 \pm 0.06$ & $16.57 \pm 1.38$ & -                & $0.55 \pm 0.12$ & $9.06 \pm 0.10$  &                       &       0.99 \\
  & core    &          & $10.94 \pm 0.04$ & $18.00 \pm 1.41$ & $1.83 \pm 0.52$  & $0.50 \pm 0.11$ & $9.02 \pm 0.10$  & 1.7                   &       1.11 \\
 \hline
 \multicolumn{10}{| c |}{Cored RCs}\\
 \hline
 NGC6503  & cusp    & 1.88     & $11.24 \pm 0.02$ & $14.07 \pm 0.85$ & -                & $0.40 \pm 0.03$ & $9.72 \pm 0.03$  &                       &       2.92 \\
   & core    &          & $11.25 \pm 0.02$ & $13.20 \pm 0.88$ & $5.14 \pm 1.35$  & $0.46 \pm 0.03$ & $9.77 \pm 0.03$  & 2.3                   &       2.72 \\
 NGC2403  & cusp    & 2.91     & $11.34 \pm 0.01$ & $15.98 \pm 0.61$ & -                & $0.32 \pm 0.03$ & $9.50 \pm 0.03$  &                       &       2.74 \\
   & core    &          & $11.32 \pm 0.01$ & $16.46 \pm 0.67$ & $0.68 \pm 0.08$  & $0.30 \pm 0.03$ & $9.48 \pm 0.04$  & 2.4                   &       2.85 \\
 F563-1   & cusp    & 3.01     & $11.47 \pm 0.12$ & $7.98 \pm 1.27$  & -                & $0.51 \pm 0.12$ & $9.02 \pm 0.10$  &                       &       1.14 \\
    & core    &          & $11.32 \pm 0.10$ & $10.77 \pm 1.65$ & $6.13 \pm 1.70$  & $0.51 \pm 0.12$ & $9.02 \pm 0.10$  & 1.1                   &       1.74 \\
 UGC05005 & cusp    & 3.31     & $11.37 \pm 0.10$ & $4.69 \pm 0.71$  & -                & $0.44 \pm 0.09$ & $9.22 \pm 0.09$  &                       &       1.03 \\
  & core    &          & $11.37 \pm 0.09$ & $5.42 \pm 0.74$  & $17.68 \pm 5.01$ & $0.48 \pm 0.11$ & $9.25 \pm 0.10$  & 1.4                   &       1.21 \\
 F574-1   & cusp    & 7.75     & $11.36 \pm 0.11$ & $7.95 \pm 1.09$  & -                & $0.54 \pm 0.12$ & $9.59 \pm 0.10$  &                       &       0.78 \\
    & core    &          & $11.13 \pm 0.08$ & $11.34 \pm 1.39$ & $6.04 \pm 1.22$  & $0.52 \pm 0.12$ & $9.57 \pm 0.10$  & 1.5                   &       1.34 \\
 NGC1090  & cusp    & 11.1     & $11.70 \pm 0.04$ & $8.90 \pm 1.35$  & -                & $0.47 \pm 0.05$ & $10.51 \pm 0.05$ &                       &       1.61 \\
   & core    &          & $11.63 \pm 0.03$ & $12.00 \pm 1.95$ & $8.42 \pm 1.47$  & $0.40 \pm 0.06$ & $10.44 \pm 0.07$ & 2.5                   &       2.30 \\
 F583-1   & cusp    & 16.53    & $11.35 \pm 0.14$ & $5.43 \pm 0.68$  & -                & $0.50 \pm 0.11$ & $8.66 \pm 0.10$  &                       &       0.70 \\
    & core    &          & $11.06 \pm 0.09$ & $8.83 \pm 0.84$  & $8.43 \pm 1.23$  & $0.50 \pm 0.11$ & $8.65 \pm 0.10$  & 1.1                   &       1.43 \\
 UGC05253 & cusp    & 24.56    & $12.09 \pm 0.01$ & $15.49 \pm 0.70$ & -                & $0.24 \pm 0.03$ & $10.85 \pm 0.04$ &                       &       3.81 \\
  & core    &          & $12.10 \pm 0.01$ & $12.64 \pm 0.74$ & $17.03 \pm 1.13$ & $0.39 \pm 0.06$ & $11.01 \pm 0.04$ & 2.4                   &       3.07 \\
 NGC5371  & cusp    & 37.23    & $12.19 \pm 0.03$ & $1.92 \pm 0.20$  & -                & $0.60 \pm 0.01$ & $11.32 \pm 0.01$ &                       &       0.37 \\
  & core    &          & $11.85 \pm 0.01$ & $29.08 \pm 1.16$ & $1.02 \pm 0.57$  & $0.14 \pm 0.01$ & $10.70 \pm 0.02$ & 2.2                   &       7.46 \\
 IC4202   & cusp    & 170.0    & $12.00 \pm 0.04$ & $18.19 \pm 0.64$ & -                & $0.53 \pm 0.05$ & $10.97 \pm 0.04$ &                       &       2.33 \\
   & core    &          & $12.01 \pm 0.02$ & $23.47 \pm 0.48$ & $8.42 \pm 0.28$  & $0.27 \pm 0.04$ & $10.69 \pm 0.05$ & 2.5                   &       2.97 \\
\hline
\end{tabular}
\end{adjustbox}
\end{table*}
Further details regarding the most reliable (75\% reliable) SPARC RCs and their fit results can be found in Table~\ref{tab:table_2} (see appendix~\ref{sec:appB} for observed RCs and their best fit mass models). In the following sections, we will discuss examples of RCs of various natures in detail.

\subsubsection{Cuspy RCs}
Among the 21 most reliable SPARC RCs, two galaxies, namely, UGC12506 and UGC02259, are unmistakably characterized as cuspy in nature with $\lbf$ values of -1.57 and -1.15, respectively. Notably, the NFW component of the best-fit cored profiles for these two RCs aligns precisely with their cuspy profile fits. In both cases, the core size ($r_c$) is either slightly less than or closely matches the radius at which the first data point is available. Additionally, the best-fit values of $c_{200}$ for UGC12506 and UGC02259 are 3$\sigma$ and 1.5$\sigma$ higher than the mean MCR, respectively, suggesting that these RCs exhibit an extremely cuspy nature.

The RCs, for which both the cusp-core distinction and the estimation of halo parameters are reliable at a 75\% level, cover a mass range of $10.9 \lesssim \lms \lesssim 12.2$. It's worth noting that, although there are only two cuspy RCs, they occupy both ends of this mass spectrum (as indicated in Table~\ref{tab:table_2}). Furthermore, when considering the RCs that are 66\% and 50\% reliable, three more cuspy RCs are found, and they predominantly belong to the lower mass range.

In essence, across the entire sample, there isn't a specific range of $M_{200}$ or $\Sigma_0$ where RCs favor a cusp profile over a core, which we can claim reliably. Consequently, with the current dataset of RCs, it isn't feasible to dismiss collision-less cold DM without offering a plausible explanation for the existence of these clearly cuspy RCs, which appear to be unaffected by the cusp-core transformation process.

\subsubsection{Cored RCs}
Among the 21 most reliable RCs, there are 10 RCs - NGC6503, NGC2403, F563-I, UGC05005, F574-I, NGC1090, F583-I, UGC05253, NGC5371, and IC4202 - which exhibit a cored nature with $\lbf > 1$. The core size ($r_c$) estimated for these RCs (excluding NGC5371) is sufficiently large, encompassing an adequate number of data points below $r_c$ to facilitate proper modeling of the inner cored section. In the case of NGC5371, the inner region (within $r < 5,\m{kpc}$) is predominantly influenced by the baryonic component ($\lsigs = 2.8$), lacking a clearly defined rising part. Both cuspy and cored profile fittings for NGC5371 fall short of accurately matching the inner data points, yet the cored profile provides a relatively better overall fit to the RC compared to the cuspy profile, especially when considering the $\Lambda$CDM prior. With the cored profile fit, the obtained $c_{200}$ for NGC5371 is 4.8$\sigma$ higher than the mean MCR, whereas the cuspy profile fit results in an even higher $c_{200}$ (5.5$\sigma$).

From figure~\ref{fig:fig_15}, it is evident that there is no specific range of $M_{200}$ or $\Sigma_0$ where the cored profile is favored. The central density of these ten cored RCs varies within the range of $\lrhobs \in (7,9.6)$. Among them, UGC05005 hosts the largest core with a size of $9.4 \pm 2.2 ,\m{kpc}$, while NGC2403 possesses the smallest core with a size of $0.63 \pm 0.07 ,\m{kpc}$. The reliability of these core parameters will be utilized in an upcoming paper to explore core properties. Given the current circumstances, the evidence supporting the existence of cores is undeniable.

\subsubsection{Model undetermined RCs}
The SPARC database contains numerous RCs for which the model is undetermined. These are the RCs for which the model selection criteria failed to provide strong evidence for either a cusp or a core, i.e., $|\lbf| < 1$. Among the 21 most reliable RCs, nine RCs fall into this category. Out of these nine galaxies, three (F568-V1, NGC0289, and UGC00128) have negative $\lbf$, and six (NGC5985, NGC5033, UGC07399, F571-V1, NGC3769, and UGC04325) have positive $\lbf$. These nine RCs could be equally well or even better fitted with both cored and cuspy profiles. However, the fit quality of either is not strong enough to favor one over the other. The core size obtained with the cored profile fit for these galaxies is not small; rather, it is significantly large, with at least two data points falling below the best-fit $r_c$.

The estimation of halo parameters for these RCs are expected to be reliabile, which is evident from the consistency in the estimated halo parameters resulting from both cuspy and cored model fits. This consistency can be observed in the fit results provided in Table~\ref{tab:table_2}. All nine of these RCs display consistent values for $\lm$ and $c_{200}$ in both the cuspy and cored profile fits. Additionally, the parameters of the stellar component ($\yd$) obtained from fitting the cuspy and cored models also align. It is worth noting that in cases where the fit quality is equal for both the cuspy and cored models, model selection tends to favor the cuspy model over the cored one, as the cored profile introduces an extra free parameter. Consequently, even though there is no explicit evidence for cores in these galaxies, one can include these nine RCs in the study of cusp-core transformation dynamics, given that cores cannot be definitively ruled out.

\subsection{CDM cosmology}
In this section, we will delve into the implications of our study for CDM cosmology, utilizing the reliable RCs from the SPARC database. We will begin by exploring the mass-concentration relation using the best-fit results from the 21 most reliable RCs. Additionally, we will touch upon the broader implications and potential future directions for CDM cosmology based on the findings of this study.

\subsubsection{Mass-Concentration relation}

\begin{figure}
    \centering
    \includegraphics[width=0.7\columnwidth]{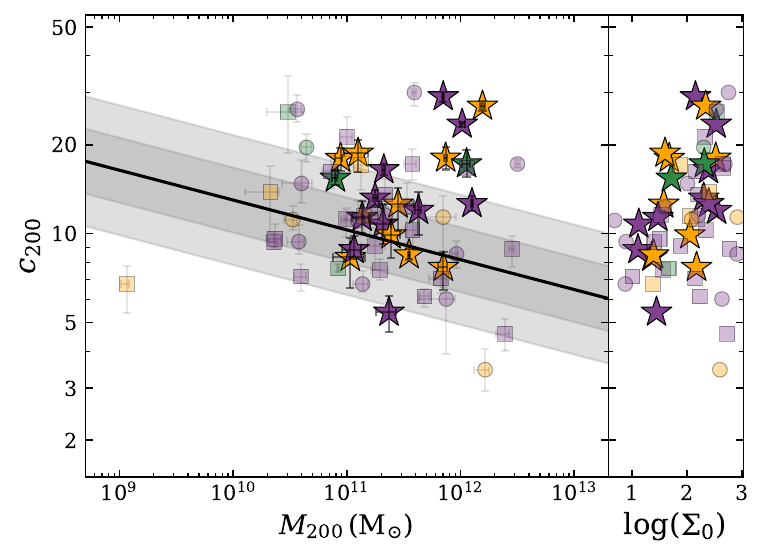}
    \caption{The best fit $c_{200}$ values are plotted as a function of $M_{200}$ (left panel) and $\lsig$ (right panel) for 75\% reliable RCs in SPARC database (star points). The results obtained for cuspy RCs are shown in green color points, those for cored RCs are shown in purple color points, and the ones for which their model was undetermined are shown using orange colored points. The same for 66\% reliable and 50\% reliable RCs are marked using squares and circles, respectively. The black line in the left panel shows the predicted MCR from \citet{Dutton_2014} along with the 1$\sigma$ (2$\sigma$) regions as dark (light) grey bands.}
    \label{fig:fig_16}
\end{figure}
Figure~\ref{fig:fig_16} illustrates the position of the 21 most reliable RCs in the Mass-Concentration Relation (MCR) plot. Cuspy RCs are denoted by green star points, cored RCs in purple stars, and model-undetermined cases in grey stars. Square points represent RCs with 66\% reliability, and circle points correspond to 50\% reliable RCs.

Even with the inclusion of the $\Lambda$CDM prior, the results reveal many galaxies with $c_{200}$ values exceeding the predicted MCR of \citet{Dutton_2014} by more than 2$\sigma$. In the right panel of figure~\ref{fig:fig_16}, the relationship between $c_{200}$ and the surface density of the stellar components, $\Sigma_0$, is depicted, which shows that for high $c_{200}$ galaxies, $\Sigma_0$ is also higher. This hints that the high concentration might be due to the influence of the baryonic components during structure formation, such as adiabatic contraction \citep{Premvijay_2023}.

\subsubsection{Future implications}
The presence of cored RCs stands as an apparent discordance to the existing theory of $\Lambda$CDM cosmology requiring a proper explanation either within the $\Lambda$CDM framework or an extension to it, to the very least. Any viable proposed solution should be able to explain the creation of central cores in galaxies of a wide range in halo mass and stellar surface density. The solution should also address the presence of cuspy RCs, which lack evidence of a significant central core. Obtaining better detailed RCs of the cuspy galaxies, such as UGC12506 and UGC02259, is important to understand why core formation failed in these galaxies. 

\section{Conclusion}
\label{sec:con}
In this paper we present a comprehensive investigation of the cusp-core problem within the context of modeling RCs using Bayesian analysis for both model selection and parameter estimation. To evaluate the capability of distinguishing between DM cores and DM cusps through RC fittings, we generated mock RCs, covering a broad spectrum of halo masses and radial extensions. We incorporated error bars and radial data points for the mock total circular velocities based on the typical characteristics of RCs found in the SPARC database. Many realizations of the mock RCs were created with varying levels of baryonic contributions and DM component scatter which underwent model fitting, similar to the procedure used for the analysis of SPARC RCs. 

We then compared the fit results to their original underlying models and parameters, allowing us to make quantitative assessments about the reliability of the inferences, for the DM halos, for which the Bayesian techniques employed here yield successful outcomes. To assess the effectiveness of the various analyses conducted on the RCs, we introduce a metric known as the reliability quotient. This parameter is calculated as the fraction of mock RCs within an ensemble of realizations that meet specific predefined validation criteria. The reliability quotient serves as an approximate measure of the likelihood or level of confidence associated with the success of the given analysis applied to the RCs in that ensemble.

Our reliability analysis points to parameter ranges in key observable quantities that one needs to consider for confidently distinguishing cusps-cores or quantifying the underlying dark matter halo. For example, we find a threshold in the stellar surface density ($\Sigma_0$), above which the reliability of distinguishing between cusps and cores diminishes. In galaxies with higher $\Sigma_0$, the central regions of the RCs are primarily influenced by the stellar velocity component. This influence has a significant impact on the small-scale features of the RCs, which, in turn, affects the ability to differentiate between cusps and cores due to the uncertainties associated with the normalizing factor, specifically the mass-to-light ratio. Consequently, previous results which suggested a preference for cusps in high mass galaxies, or vice-versa, should be interpreted as the failure of the model selection criteria employed to distinguish cuspy-vs-cored profiles in high $\Sigma_0$ galaxies. Interestingly, the threshold value of $\lsigs = 2.5$ inferred in our work agrees with the ``magic'' value at which such cusp-core dichotomy was observed in previous works (for e.g., \cite{Zentner_2022},\cite{Frosst_2022}).

Additionally, we analyzed the success of estimating both halo and core parameters. We show that the radial extent of the RCs plays a pivotal role in determining the reliability of obtaining halo parameters with high levels of precision and accuracy. While RCs of low mass galaxies (in virtue of the low stellar surface density) excel at reliably distinguishing between cusps and cores, the existing quality of RC data falls short when it comes to making reliable claims regarding their associated halo parameters. Furthermore, the reliability of estimating core parameters shows a decline when dealing with RCs of more massive haloes and limited radial extensions. This nuanced interplay between various factors highlights the intricate nature of parameter estimation within the context of RC analysis.

Applying our formalism to the RCs in SPARC dataset, we identified 21 RCs that falls within the region where reliable model selection and halo parameter estimation can be made robustly, with at least a 75\% success rate. Of this subset, 2 RCs exhibit distinctive cuspy features, while 10 are classified as cored systems. However, for the remaining 9 RCs, the true model classification remains undetermined, preventing a clear distinction between cusp or core. Notably, within the narrow range of halo mass encompassed by the majority of these reliable RCs, no significant classification could be found where either cusp or core is observed predominantly.

An estimation of the mass-concentration relation, using the 21 most reliable RCs, reveals that $\approx 15$\% of this sample is more than 3$\sigma$ away from what is expected from numerical, DM only, simulations. 
This departure underscores the potential influence of baryonic components on the concentration of the dark halo and connects galaxy formations within their parent dark matter halos. 

The formalism introduced in this paper is sufficiently general to be applied to a variety of datasets and problems. For DM studies using RCs, our formalism can be used to get a subset of RCs, i.e gold samples, with varying degrees of reliability. We have used such selective RC's in a number of dark matter studies, for example, issues of diversity and peculiarities in RCs \citep{MM_2024b, MM_2024c}, the implications for direct DM detection experiment \citep{MM_2024a} and phase-space of DM in nearby observed galaxies \citep{MM_2023}.

\appendix

\section{Choice of priors}
\label{sec:appA}
\begin{figure}
    \centering
    \includegraphics[width = 0.5\linewidth]{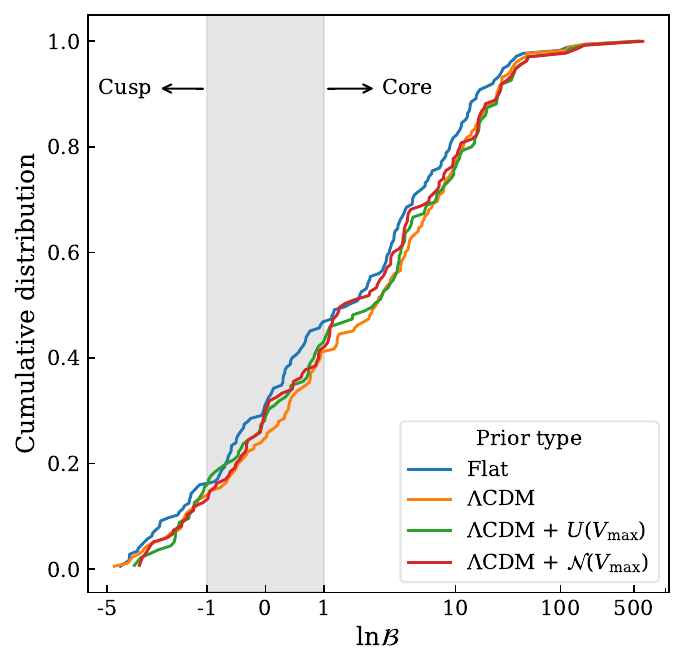}
    \caption{CDF of $\lbf$ obtained with various set of priors. $U$ represents uniform prior and $\mathcal{N}$ represents Gaussian prior.}
    \label{fig:fig_A1}
\end{figure}
We have implemented various choices of priors on DM parameters to analyze the RCs. The various priors implemented and the resulting cumulative distribution (CDF) of the Bayes factors are plotted in figure~\ref{fig:fig_A1}. The four types of priors considered here are: (i) flat prior on $c_{200}$, (ii) $\Lambda$CDM prior as log-normal distribution on $c_{200}$, (iii) $\Lambda$CDM prior on $c_{200}$ along with regularization prior on $\vmax/V_{\m{flat}}$ as uniform prior between $1/\sqrt{2},\sqrt{2}$ and, (iv) $\Lambda$CDM prior on $c_{200}$ along with regularization prior as Gaussian distribution on $\vmax$ around $V_{\m{flat}}$ as discussed in Section~\ref{ssec:bayes}.

While implementing regularization prior as simple flat prior helped to break the degeneracy in DM parameters of many RCs, it also introduced extra degeneracies in many RCs which preferred $\vmax/V_{\flat}$ outside of the range $[1/\sqrt{2},\sqrt{2}]$. Extending the range of flat regularization prior to [1/2,2] reduced the number of RCs that were affected. In order to mitigate this issue, we implemented the Gaussian prior on $\m{log}({\vmax/V_{\m{flat}}})/\m{log}(2)$ centered at 0 and with a standard deviation of 0.5. 

\begin{figure}
    \centering
    \includegraphics[width = 0.5\linewidth]{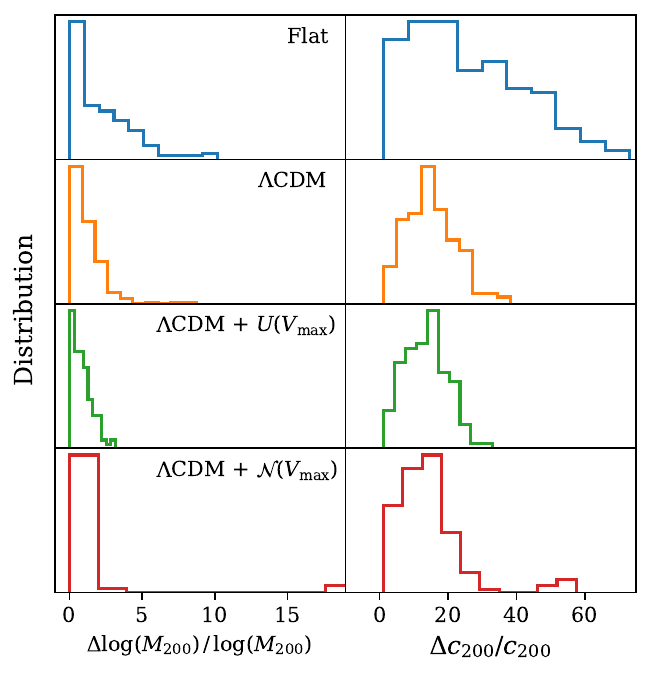}
    \caption{Distribution of the percentage errors of $\lm$ and $c_{200}$ obtained for various sets of priors. $U$ represents uniform prior and $\mathcal{N}$ represents Gaussian prior.}
    \label{fig:fig_A2}
\end{figure}
The effectiveness of introducing a regularization prior, specifically a Gaussian prior on $\vmax$, becomes evident upon reviewing the results illustrated in figure~\ref{fig:fig_A2}. The distributions of the percentage errors for both $\lm$ and $c_{200}$ obtained with various sets of priors are presented, highlighting the impact of this regularization approach. A noteworthy observation is the reduction in the standard deviation of the marginalized posteriors when an additional prior is introduced, this happens provided the additional prior agrees with the posteriors obtained without it. From figure~\ref{fig:fig_A2}, it is apparent that the percentage errors (at the 67\% confidence level) are minimized when implementing $\Lambda$CDM prior along with the Gaussian regularization prior. Furthermore, this error reduction is substantial when compared to the case with a flat prior.

\section{Reliable RCs}
\label{sec:appB}
\begin{figure}
    \centering
    \includegraphics[width = 0.6\textwidth]{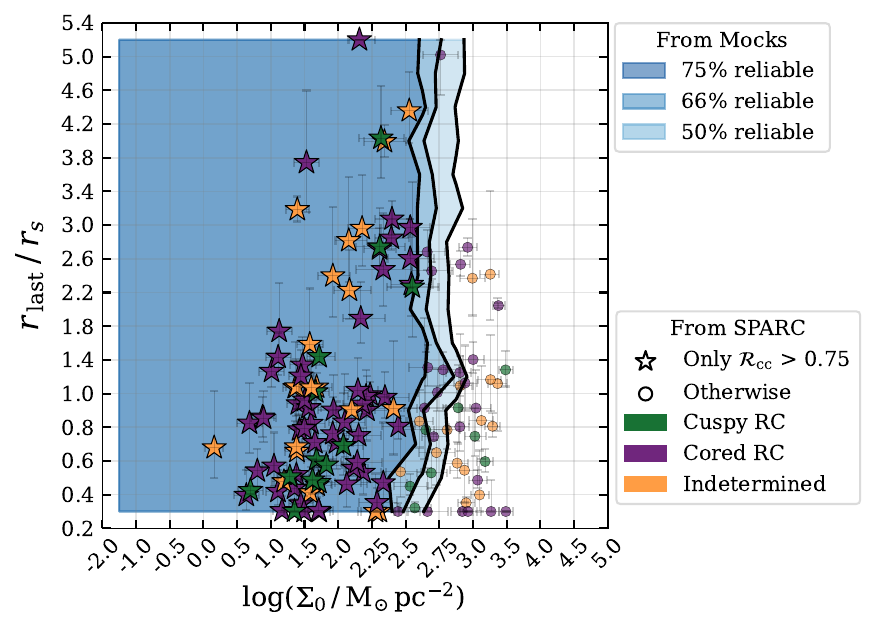}
    \caption{Position of SPARC galaxies on the reliability space of cusp-core distinction, irrespective of comparable reliability in halo-parameters. Galaxies that are  $>75$\% reliable are marked using stars and all others with circles. The color of the data points indicates the inferred cusp/core based on Bayes factor, where green (purple) color represents those which prefer cuspy (cored) profile. Orange colored points are for those RCs in which Bayes factor criteria for model selection were inconclusive.}
    \label{fig:fig_B1}
\end{figure}

\begin{figure}
    \centering
    \includegraphics[width = 0.6\textwidth]{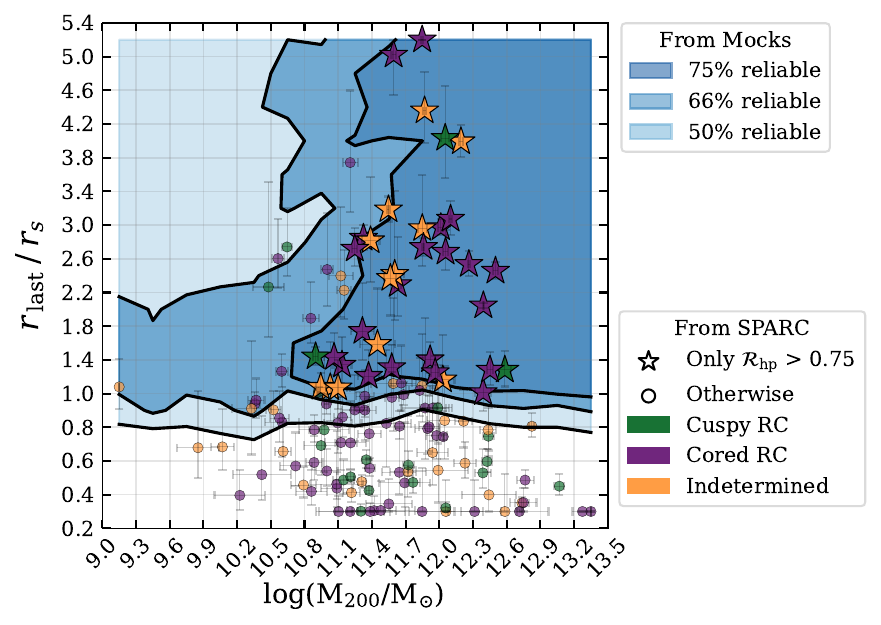}
    \caption{Position of SPARC galaxies on the reliability space of halo-parameters estimation, irrespective of reliability of cusp-core distinction. The marker and color schemes are same as in figure~\ref{fig:fig_B1}.}
    \label{fig:fig_B2}
\end{figure}

Figures~\ref{fig:fig_B1} and \ref{fig:fig_B2} shows the position of SPARC galaxies on the reliability space of cusp-core distinction and halo-parameter estimation, respectively. Those falling in the 75\% reliable region of respective plots are marked using stars and circle are used for those falling outside. Nearly 34\% (73\%) of the galaxies fall outside the 75\% reliable region for cusp-core distinction (halo-parameter estimation). (We also note that, out of the 21 RCs (figure~\ref{fig:fig_15}) that lie in the 75\% reliable region of both cusp-core and halo-parameter estimations, 4 of them has a bulge component.)

\begin{figure*}
    \centering
    \includegraphics[scale = 0.7,draft = False]{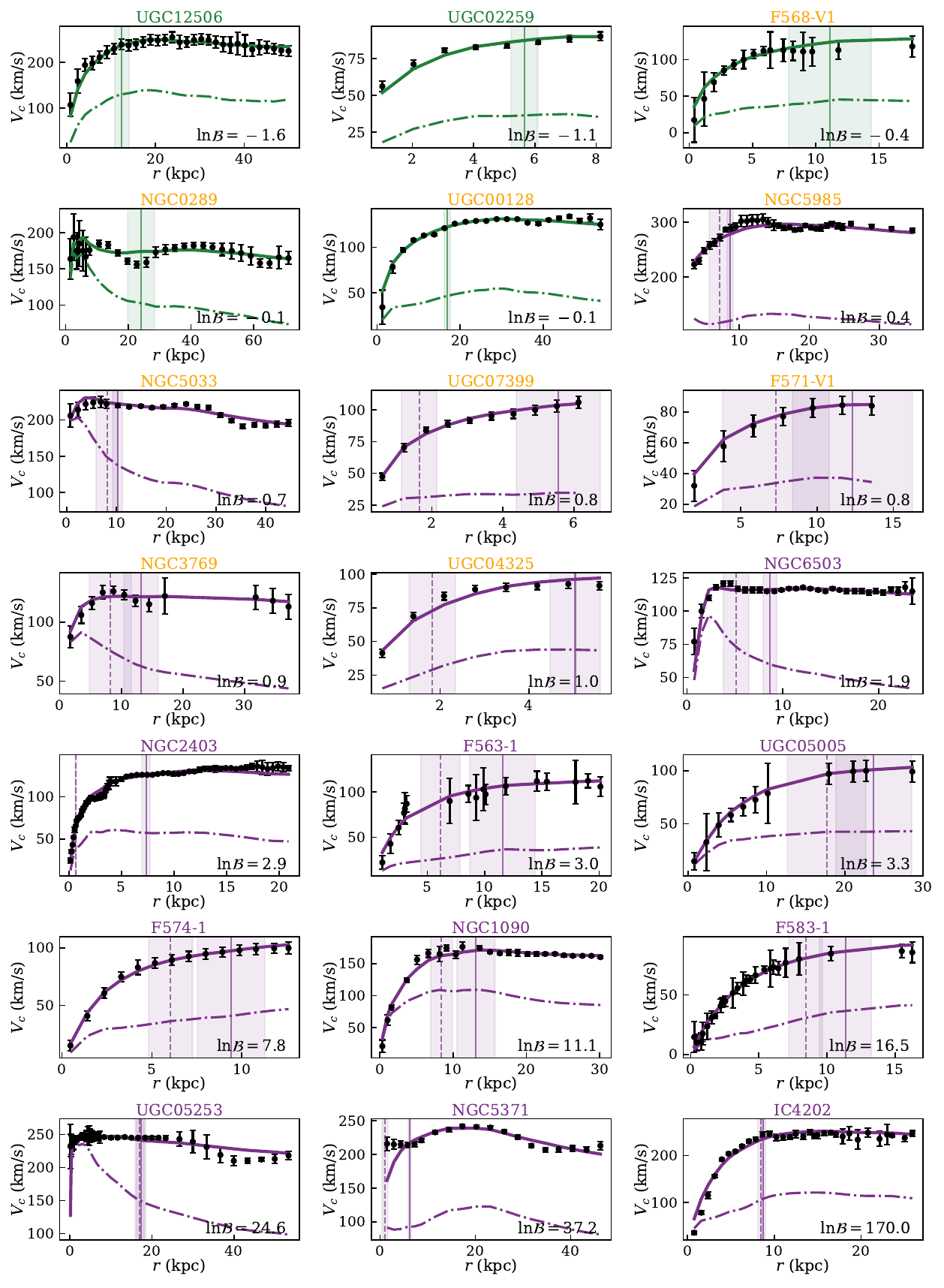}
    \caption{75\% reliable RCs in SPARC database, where cusp-core distinction and halo parameter estimation are the most reliable. See the text for the plot description.}
    \label{fig:fig_rel75}
\end{figure*}

\begin{figure*}
    \centering
    \includegraphics[scale = 0.7,draft = False]{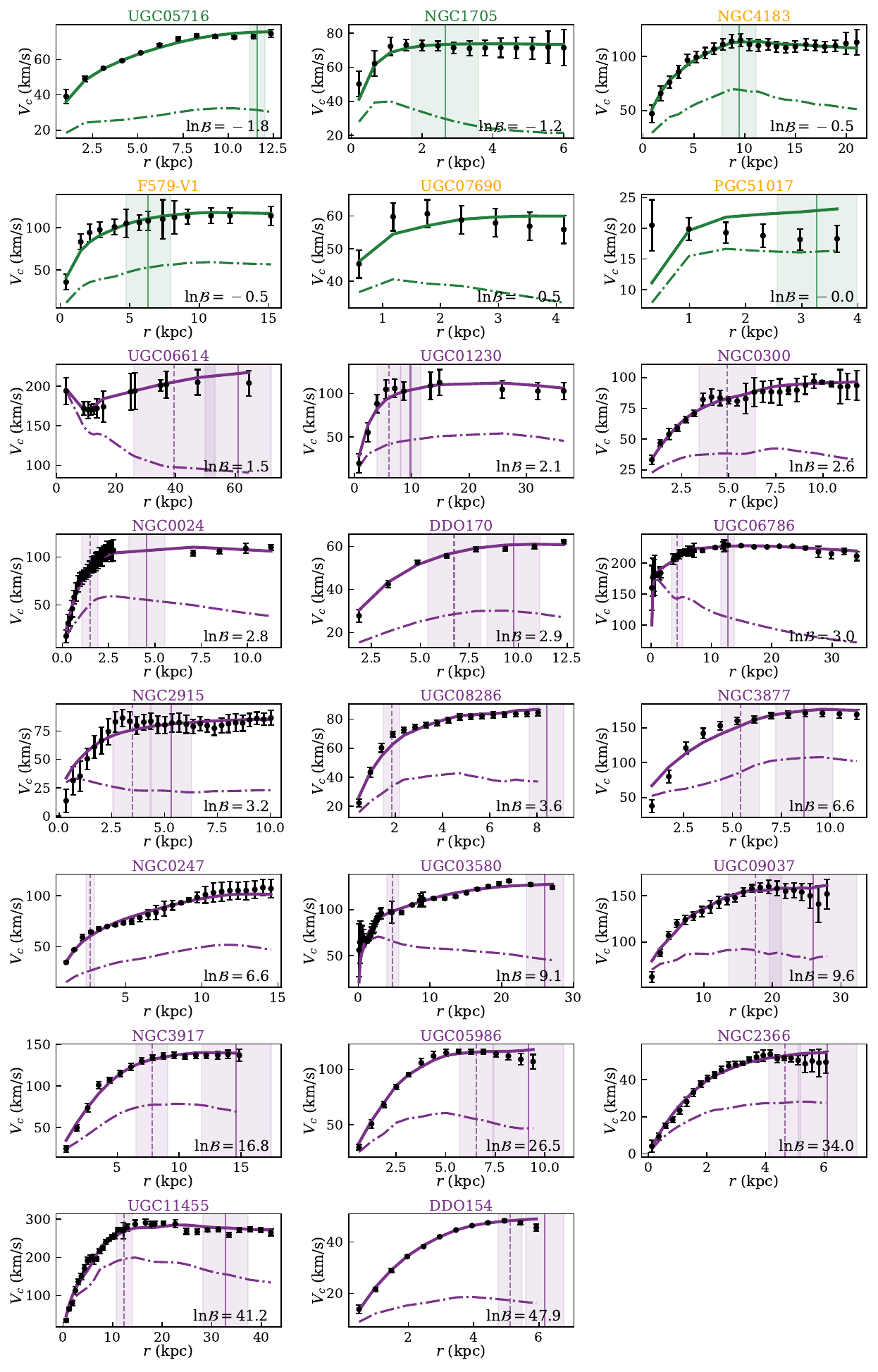}
    \caption{66\% reliable RCs in SPARC database (see text for the plot description).}
    \label{fig:fig_rel66}
\end{figure*}

\begin{figure*}
    \centering
    \includegraphics[scale = 0.7,draft = False]{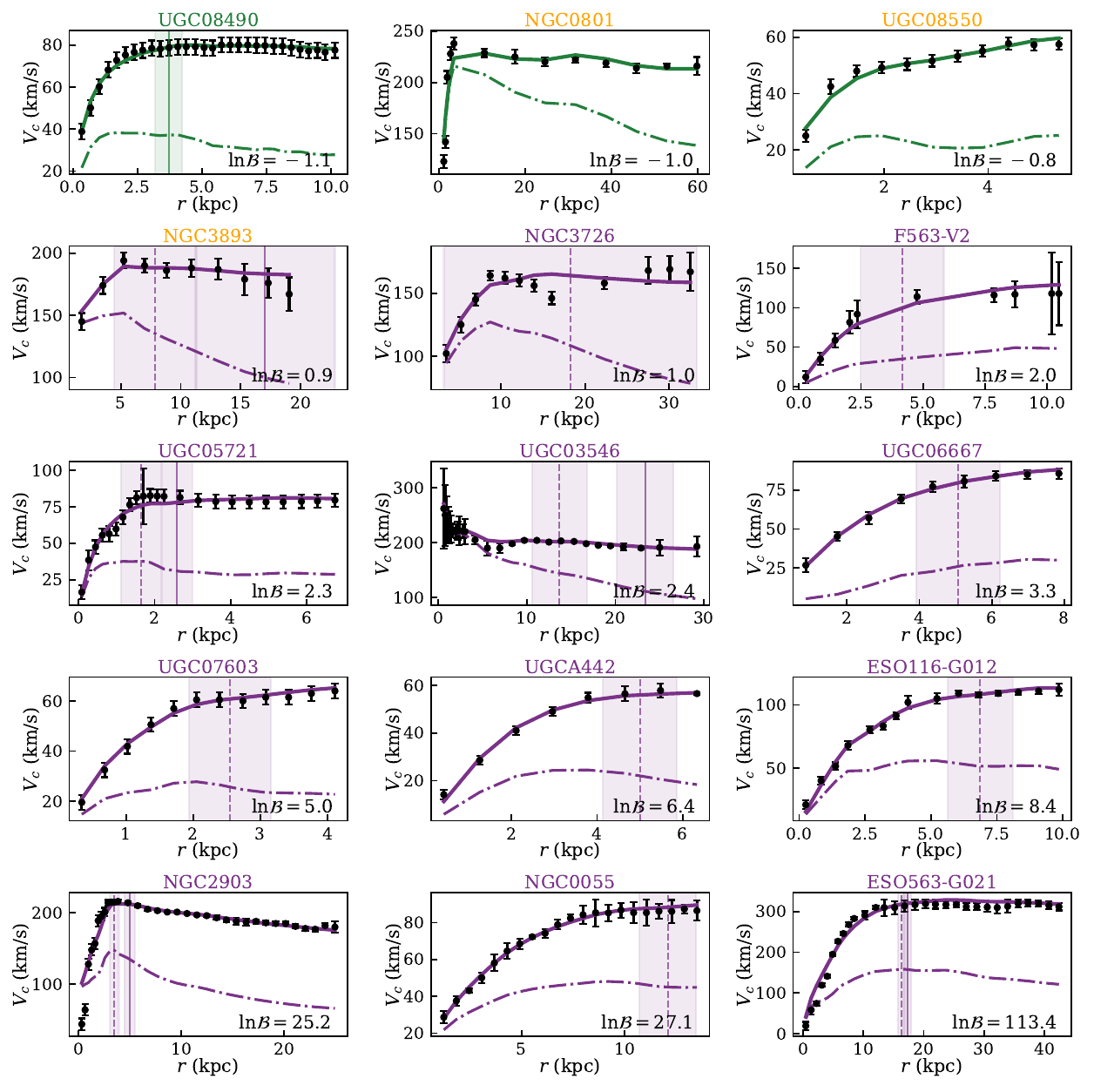}
    \caption{50\% reliable RCs in SPARC database (see text for the plot description).}
    \label{fig:fig_rel50}
\end{figure*}
The RCs and their fit results for various sets of reliable samples in the SPARC database are shown in Figs.~\ref{fig:fig_rel75}, \ref{fig:fig_rel66} and, \ref{fig:fig_rel50} with 75\%, 66\% and 50\% reliability for cusp-core distinction and estimation of halo parameters, respectively. The name of the galaxy of each RC is noted as the title of each panel, where the color used for the title represents the best fit model of that RC. Green color represents cuspy RC, purple represents cored RCs and orange for RCs where the model selection criteria failed to identify either cusp or core as the best fit model. The observed data are plotted in black error bars. The best fit total velocity curve is plotted in solid line, where the green line is used if the best fit model is cuspy ($\lbf < 0$) in nature and purple if its cored ($\lbf$  > 0). The solid horizontal line shows the best fit $r_s$ and dotted purple lines show the best fit $r_c$ along with their 1$\sigma$ regions shaded.

\begin{acknowledgments}
The authors would like to thank Aseem Paranjape and Joe Silk for constructive discussions. MM would like to thank Aakash Pandey for discussions during the initial stage of this project. MM sincerely wishes to thank TIFR for its hospitality, funding, computational facilities, and especially DTP for giving him the opportunity to visit TIFR for multiple long durations during which this work was completed. MM would like to thank Rishi Khatri for financial support which made many TIFR visits possible. SM acknowledges the support of the Department of Atomic Energy, Government of India, under project no. 12-R\&D-TFR- 5.02-0200.
\end{acknowledgments}

\newpage
\bibliographystyle{unsrtnat}
\bibliography{references}

\end{document}